# DESIGN AND IMPLEMENTATION OF A DIFFERENTIATED SERVICE BASED QOS MODEL FOR REAL-TIME INTERACTIVE TRAFFIC ON CONSTRAINED BANDWIDTH IP NETWORKS

*Thesis submitted to*

*Indian Institute of Technology, Kharagpur*

*for the award of the degree*

*of*

**Master of Science**

*by*

**Sruti Gan Chaudhuri**

*under the guidance of*

**Dr. C.S. Kumar**
**Department of Mechanical Engineering**

*and*

**Prof. R.V. Raja Kumar**
**Department of Electronics and Electrical Communication Engineering**

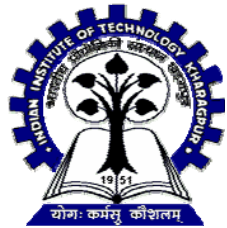

**G. S. SANYAL SCHOOL OF TELECOMMUNICATIONS**

**INDIAN INSTITUTE OF TECHNOLOGY, KHARAGPUR, INDIA.**

**February 2010**

# APPROVAL OF THE VIVA-VOCE BOARD

/02/2010

Certified that the thesis entitled DESIGN AND IMPLEMENTATION OF A DIFFERENTIATED SERVICE BASED QOS MODEL FOR REAL-TIME INTERACTIVE TRAFFIC ON CONSTRAINED BANDWIDTH IP NETWORKS submitted by SRUTI GAN CHAUDHURI to the Indian Institute of Technology, Kharagpur, for the award of the degree Master of Science has been accepted by the external examiners and that the student has successfully defended the thesis in the viva-voce examination held today.

(Member of the DSC)        (Member of the DSC)        (Member of the DSC)

(Supervisor)                                           (Supervisor)

(External Examiner)                                    (Chairman)

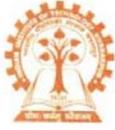

**INDIAN INSTITUTE OF TECHNOLOGY KHARAGPUR-721302**

भारतीय प्रौद्योगिकी संस्थान खड़गपुर

**G. S. SANYAL SCHOOL OF TELECOMMUNICATIONS**

/02/2010

# CERTIFICATE

This is to certify that the thesis entitled **Design and Implementation of a Differentiated Service based QoS Model for Real-Time Interactive Traffic on Constrained Bandwidth IP Networks,** submitted by **Sruti Gan Chaudhuri** to Indian Institute of Technology, Kharagpur, is a record of bona fide research work under our supervision and we consider it worthy of consideration for the award of the degree of Master of Science of the Institute.

Signature of Supervisor(s)

| | |
|---|---|
| **Prof. R.V. Raja Kumar** | **Prof. C S. Kumar** |
| Department of Electronics and Electrical Communication Engineering | Department of Mechanical Engineering |
| Indian Institute of Technology, | Indian Institute of Technology, |
| Kharagpur, India. | Kharagpur, India. |

Phone: (91/0-3222) 282265    Gram : Technology Kharagpur    Fax : (91/0-3222) 282266

iii

# Acknowledgements

I would like to take this opportunity to make an attempt to express my gratitude towards all those people who have played an important role in some way or another to help me achieve whatever little I have.

First and foremost, I wish to express my sincerest respects to my supervisors Prof. C.S Kumar of Department of Mechanical Engineering, IIT Kharagpur and Prof. R.V. Raja Kumar of Department of Electronics and Electrical Communication Engineering, IIT Kharagpur. This work would not be in the form it is today without their intellectual inputs and advices throughout the duration of my Master's program. I am also grateful to Prof. Saswata Chakraborty from G.S.Sannyal School of Telecommunication, IIT Kharagpur and Prof. S.S Pathak of Department of Electronics and Electrical Communication Engineering, IIT Kharagpur, for giving me valuable suggestions time to time.

I wish to thank my lab seniors (Prashant, Umesh, Madhav, Sarath, Sambit), lab mates (Lavanya, Shrilekha, Shupi, Kausani, Vipin, Navraj) and lab staffs (Mihirda and Kaushikda) for providing me a comfortable and enjoyable working environment. I am also obliged to all my friends who have always supported me directly or indirectly, to continue my work. Last but not the least I would like to express my regards to my parents for always believing in me and motivating me to do my best.

# DECLARATION

I certify that
    a. The work contained in the thesis is original and has been done by myself under the general supervision of my supervisor(s).
    b. The work has not been submitted to any other Institute for any degree or diploma.
    c. I have followed the guidelines provided by the Institute in writing the thesis.
    d. I have conformed to the norms and guidelines given in the Ethical Code of Conduct of the Institute.
    e. Whenever I have used materials (data, theoretical analysis, and text) from other sources, I have given due credit to them by citing them in the text of the thesis and giving their details in the references.
    f. Whenever I have quoted written materials from other sources, I have put them under quotation marks and given due credit to the sources by citing them and giving required details in the references.

<div style="text-align: right;">Signature of the Student</div>

# Brief Biography of the Author

**Miss Sruti Gan Chaudhuri** is from **Kolkata**, West Bengal. She passed **Madhyamik (10)** from West Bengal Board of Secondary Education (**WBBSE**) in **1998** and **Higher Secondary (10+2)** from West Bengal Council of Higher Secondary Education (**WBCHSE**) in **2000**. She did **Bachelor of Technology (Btech)** in **Computer Science and Engineering** in **2004** from **University of Kalyani, West Bengal.**

After that she joined **Indian Institute of Technology, Kharagpur**, as a **Junior Project Assistant (JPA)** in a project entitled **"Establishment of Nation Wide Quality of Service Network Test Bed "**, sponsored by **Ministry of Information Technology, ERNET Delhi**, from **October 2004 to July 2007**. During this period she joined to the **MS (Master of Science by research)** program at **G. S. Sanyal School of Telecommunications, IIT Kharagpur.** She worked as **Senior Project Assistant (SPA)** in the same project from **August 2007 to April 2008**.

Currently she is working as a **Project Linked Personnel (PLP)** in project on **"Localization and Routing in Wireless Sensor Networks"**, at **Advanced Computing and Microelectronics Unit of Indian Statistical Institute, Kolkata** from **May 2008 to till date.**

# Dedication

To My Parents.

# List of Abbreviations

| | |
|---|---|
| AF | Assured Forwarding |
| ATM | Asynchronous Transfer Mode |
| BE | Best Effort |
| CBR | Constant Bit Rate |
| CoS | Class of Service |
| DCCP | Datagram Congestion Control Protocol |
| DiffServ | Differentiated Service |
| DITG | Distributed Internet Traffic Generator |
| DSCP | Differentiated Service Code Point |
| EF | Expedited Forwarding |
| FIFO | First In First Out |
| ICMP | Internet Control Message Protocol |
| IDT | Inter Departure Time |
| IEEE | Institute of Electrical and Electronics Engineers |
| IETF | Internet Engineering Task Force |
| IntServ | Integrated Service |
| IP | Internet Protocol |
| ITU | International Telecommunication Union |
| JMF | Java Media Framework |
| JPEG | Joint Photographic Expert Group |
| LBNL | Lawrence Berkeley National Laboratory |
| MPEG | Motion Picture Experts Group |
| MPLS | Multi Protocol Level Switching |
| NS2 | Network Simulator 2 |
| NTP | Network Transmission Protocol |
| OSPF | Open Shortest Path First |
| OTcl | Object oriented Tcl |
| PHB | Per Hop Behavior |
| PS | Priority Scheduling |
| QoS | Quality of Service |
| RED | Random Early Detection |
| RFC | Request For Comments |
| RR | Round Robin |
| RSVP | Resource reSerVation Protocol |
| RTCP | Real time Transmission Control Protocol |
| RTP | Real time Transmission Protocol |
| SCTP | Stream Control Transmission Protocol |
| TCP | Transmission Control Protocol |
| ToS | Type of Service |
| UCB | University of California, Berkeley |
| UDP | User Datagram Protocol |
| VBR | Variable Bit Rate |
| VLAN | Virtual Local Area Network |
| VoIP | Voice Over IP |
| WAN | Wide Area Network |
| WRR | Weighted Round Robin |

# List of Figures

# List of Tables

# Abstract

The traditional service model for internet protocol (IP) provides only the Best Effort (BE) service where the same priorities are assigned to both real-time and non-real-time traffic. The BE service model does not guarantee Quality of Service (QoS) to real-time traffic. The real-time interactive applications need to be supported with tight guarantees of QoS.

In this thesis work, a QoS model for real-time interactive traffic on a real network with constrained bandwidth and real-time traffic has been proposed. The model supports tight guarantees of QoS to real-time interactive traffic without over provisioning of bandwidth. A dynamic scheduling model which is adaptive to input data rate of traffic has been proposed. In this model, A Differentiated Service (DiffServ) based approach is proposed for QoS provisioning. The packets are classified and distributed among finite number of queues with limited buffer based on different priorities and total available bandwidth. The model proposes a mechanism to derive the weighted service rates and queue length distribution so as to meet the requirement of low packet loss and delay for real time interactive traffic in the QoS engineered network. An adaptive queuing strategy is proposed so that minimum bandwidth in used for real time traffic. This ensures maximizing availability to best effort traffic. The model assumes constrained bandwidth without having to over provision the network resources and thus keeping the cost low. A modified version suitable for testing on a real network is also presented.

Experimental verification of these in a test bed network in a laboratory as well as on a real network has been carried out. The results of the QoS provisioning model for different sources of real-time traffic such as video conferencing equipment, robotic surveillance camera has also been shown.

The thesis also introduces a real-time Variable Bit Rate (VBR) traffic tuning parameter for controlling the service of VBR traffic to give better and fair performance to rest of the traffic.

# Keywords

# Contents

# Chapter 1
# Introduction

The internet applications are rapidly shifting to triple play service oriented usages. There is an increasing need to engineer Quality of Service (QoS) in a network to support the service requirement of the real-time interactive traffic, such as, video conferencing, IP telephony, online gaming, e-commerce etc. The standard best-effort service model supported by the internet, works very well with conventional data applications like file transfer, electronic mail and web browsing. But it does not guarantee timely and actual delivery of packets. Multimedia applications with real-time properties require tight guarantees in terms of end-to-end packet delay and packet loss.

In order to meet these requirements and to manage the multitude of interactive real-time applications, QoS has to be suitably implemented in the various network devices, such as switches and routers. The limited bandwidth and resource availability in many data networks act as a constraint in deploying QoS.

Initial efforts on providing QoS for IP networks focused on the Integrated Services (IntServ) model [1]. This model relied on a flow-based, connection-oriented approach to deliver QoS for IP networks. The IntServ model is matured and a good amount of work has been completed. However, it has not been widely deployed due to a variety of concerns. Here the primary issue is scalability of bandwidth. With IntServ, intermediate routers need to save per-flow state information. Another concern is the end-to-end connection-oriented approach of Resource reSerVation Protocol (RSVP) [2]. The above issues result in an architecture that is complex to implement and deploy. This is the most important reason why the Differentiated Services initiative has started.

The Differentiated Services (DiffServ) Architecture [3] attempts to move the complexity to the edge of the network and keep the core simple. It intends to provide different levels of service in the internet without the need for per-flow state and signaling at each router. The framework includes the following elements:

1. Traffic conditioning element at the edge of the network that marks packets to receive certain levels of service at the backbone, polices packets and performs traffic shaping to ensure that packets entering the backbone conform to network policies.

2. Edge routers that treat packets differently depending on the packet marking completed by the edge device.

3. Allocation or policy mechanism that translates into end-to-end QoS levels of service seen by the end-users. Traditionally DiffServ assumes the availability of sufficiently large bandwidth and resources of network devices as an over provisioned network.

A lot of research work has been done and also continuing to provide good QoS to real-time traffic. Following are some QoS implementation models presented in the literature.

## 1.1 Literature survey

The DiffServ model is aimed at supporting service differentiation for aggregated traffic in a scalable manner. Many approaches have been proposed to realize this model. The absolute DiffServ mechanism [4], [5], [6] provides IntServ-type end-to-end absolute performance guarantees without per-flow state in the network core. On the other hand relative DiffServ mechanism provides per-hop, per-class relative services. Relative DiffServ approach [7] can not provide end to end guarantees. Instead, each router only guarantees that the service invariant is locally maintained, even though the absolute end-to-end service might vary with networking conditions.

Many real-time applications, such as VoIP, military command and control systems or industrial control systems, demand a predefined end-to-end deadline. Packets received beyond these end-to-end deadlines are considered useless. The real-time applications rely on absolute differentiated services in order to have guarantee on the end-to-end delay.

Nichols et al. [4] proposed a premium service model, which provides the equivalent of a dedicated link between two access routers. It provides absolute DiffServ in priority driven scheduling networks with two priorities, in which high priority is reserved for premium services. R. Cruz [8] proposed an algorithm that provides both guaranteed and statistical rate and delay bounds and addresses scalability through traffic aggregation and statistical multiplexing. Wang et al. [9] proposed a methodology for providing absolute DiffServ for real-time applications in networks that use static-priority schedulers.

B. Nandy at el. [10] proposed a connection less framework to provide QoS in IP network. This model is based on IntServ, but does not use RSVP. It addresses scalability concerns of IntServ by removing the need for a connection-oriented reservation setup mechanism and replaces it with a DiffServ-like mechanism to consistently allocate bandwidth, end-to-end in a network. This approach allows the device to automatically detect QoS requirements of the applications without the need for application-level signaling. A priority-based scheduling mechanism is used with a variant of weighted round-robin.

F. Palmieri [11] presents an approach to implement end-to-end QoS according to the DiffServ based model to achieve significant improvements in the performance of high traffic multimedia streaming networks. This model uses an intelligent edge device to accurately classify and manage traffic based on pre-determined policies, and marks the traffic appropriately before it enters the backbone network. DiffServ enabled core routers can then switch the traffic quickly and efficiently, providing true end-to-end quality of service for multimedia streams. The model uses bandwidth allocation scheme via the CAR-based policing feature for carefully engineering the network capacity to meet bandwidth commitments during periods of congestion.

T. Ahmed at el. [12] presents a practical approach to manage multimedia traffic in a DiffServ network, using network monitoring, feedback and control. Bandwidth monitors are installed in each node of DiffServ domain and interact with a policy server. Depending on the network state, the policy server decides the policies of the traffic. The implementation of the selected policies typically leads to accepting, remarking, or dropping the multimedia traffic entering the network.

I. McDonald [13] proposed a model where information about the priority of packets and expiry times are used by the transport layer to reorder or discard packets to optimize the use of the network. This can be used for video conferencing to prioritize important data. This algorithm is implemented as interface to the Datagram Congestion Control Protocol (DCCP) and it gives better improvements to video conferencing using standard UDP and TCP.

H. Wu at el. [14] proposed a soft DiffServ based QoS architecture enabled communication approach to the service providers to offer each customer a range of services that are differentiated on the basis of performance. This paper presents a new enhanced communication approach at the end system to support distributed multimedia applications. Two new mechanisms are highlighted in this approach: enhanced communication stack to support differentiation within one flow and the network-awareness provision for applications at the end system.

In this thesis work, the scalable architecture Differentiated Service has been considered with constrained bandwidth bottleneck link without over provisioning of bandwidth at the access side and a QoS scheduling model have been proposed to guarantee the performance for real-time interactive traffic. A QoS scheduling algorithm which is adaptive to the input data rates of different streams of traffic has been proposed to fairly serve the different real-time traffic as well as the other non real-time traffic.

## 1.2 Motivation of the thesis work

Recent trends indicate that the internet is being rapidly flooded by multiple traffic flows (e.g. video traffic from you-tube, skype traffic, IP TV etc.) consuming a considerable amount of bandwidth from the network. Such traffic requires a good management of the network and not only just providing of additional bandwidth by a service provider, but also the routers need to implement QoS management and control techniques. From administration point of view the QoS implementation is shifting towards the edge of the network to make the backbone of the network simple and improving the scalability of QoS implementation. The network with limited resources such as bandwidth at the edge and limited buffer memory of the router pose challenges to the implementation of QoS in a scalable manner for real-time interactive applications. The edge network access speed is often low and of the order of 2 Mbps. The edge routers are deployed with small buffer memory (typically four queues with 512 Kb) that limits the number of simultaneous schedulers for various traffic types. Planning a QoS implementation with these factors requires a judicious allocation of resources based on DiffServ and the volume of the traffic is allowed by a call admission control scheme without over provisioning the bandwidth. In this thesis work, the problem of resource allocation with the above constraints is addressed. It is presumed that the service growth of the network traffic can be managed by progressively relaxing the constraints without over provisioning (e.g. raising bandwidth to 8 Mbps and/or adding routers with 8 queues and more memory etc.).

## 1.3 Objective of the thesis work

Over provisioning of resources for a network is expensive. Hence, a good model for QoS management will be well suited for a network with a constrained bandwidth. Therefore, the objective of this thesis work is formulated with the following activities:

1. To model a Differentiated services based QoS strategy for real-time interactive traffic at the edge of the network with constrained bandwidth bottleneck link.

2. To develop and deploy a proposed QoS model based IP network test-bed in the laboratory to monitor the performance of real-time interactive traffic across this test-bed under constrained bandwidth bottleneck link condition.

## 1.4 Organization of the thesis

The thesis consists of 6 chapters. Chapter 1 presents the introduction, literature review and objective of this thesis work. Chapters 2 and 3 give the overview of real-time traffic and QoS in real-time communication respectively.

Chapters 4 and 5 are the core chapters of the thesis. Chapter 4 presents description of the proposed QoS model in detail. The simulations of the proposed model and simulated results also have been discussed in this chapter. Chapter 5 presents the experimental study and analysis of the proposed QoS model on a test-bed setup in the laboratory. The description of the experimental setup at the laboratory has been described in this chapter.

Finally chapter 6 comes with the conclusions of the thesis. It summarizes the work done in the thesis and presents the contribution and future work of this thesis. The references and appendixes are also attached at the end of the thesis. A brief description of the modeling in network simulator (NS2) used to implement DiffServ QoS, the traffic generator (DITG) used in the experimental study to generate various types of traffic and the QoS configuration of the router used for the experimental study have been presented in appendix I, II and III respectively.

# Chapter 2
# Real-Time Traffic in Packet Data Networks

In this chapter an overview of real-time transmission is presented. A protocol support other than conventional TCP and UDP required for real-time transmission. This protocol has been described in this chapter. At the end of the chapter different types of real-time traffic are introduced briefly.

## 2.1 Introduction to real-time transmission and supporting protocol

Internet Protocol is the heart of modern networks but by design it is unreliable i.e. it does not ensure that all packets sent from one endpoint arrive successfully at the other end point. Data applications require a reliable transport. To ensure this, networks use Transmission Control Protocol (TCP) on top of IP (TCP/IP). TCP ensures that, each packet that is sent arrives at the other end, and it resends a packet if one is lost. The TCP protocol uses packet loss as its flow control mechanism. Data applications are very busty in their utilization of the network bandwidth. Figure 2.1 is a graph of a typical data communication application, showing periods of low and high network utilization [17].

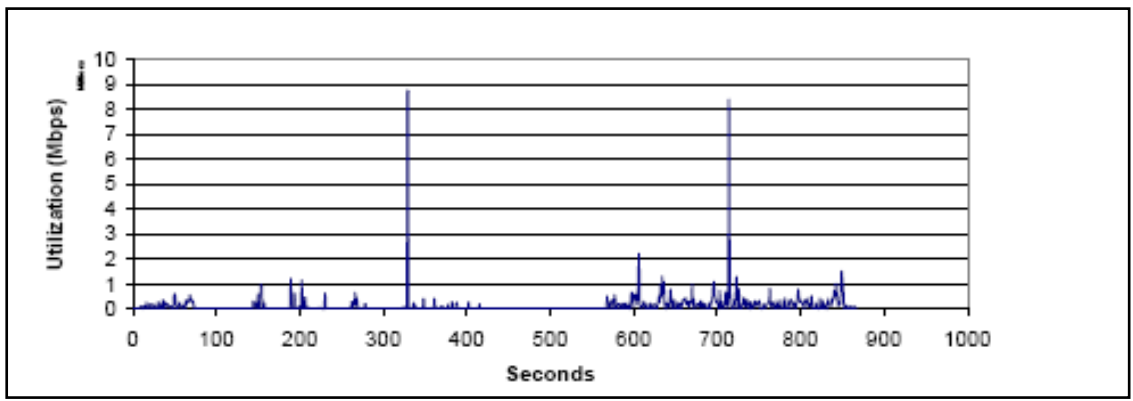

Figure 2.1: Typical data bandwidth utilization

Real time media streams implies the sending or receiving of stored or live media (voice or video) broadcast or videoconferencing over the internet. Real-time traffic is very different in its characteristics. It results from the output of a codec which is sampling a continuous real-world environment (speech or images) and transmitting constant updates of this information to reproduce the image or speech. So the bandwidth utilization of voice and video is sustained during the time the application is running. Figure 2.2 is a graph of a 384K video conference, showing both the audio and video streams and their relatively constant use of bandwidth during operation [17].

The real-time streams are delay sensitive. It samples and reproduces a continuous event, such as speech or image. Individual data samples must arrive at the destination end to be played at the right time. If a packet is late, or is lost in transit, then there will be a gap in the information available to the player, and the quality of the audio or video reproduction will degrade. This degradation is significant with increasing delay or loss.

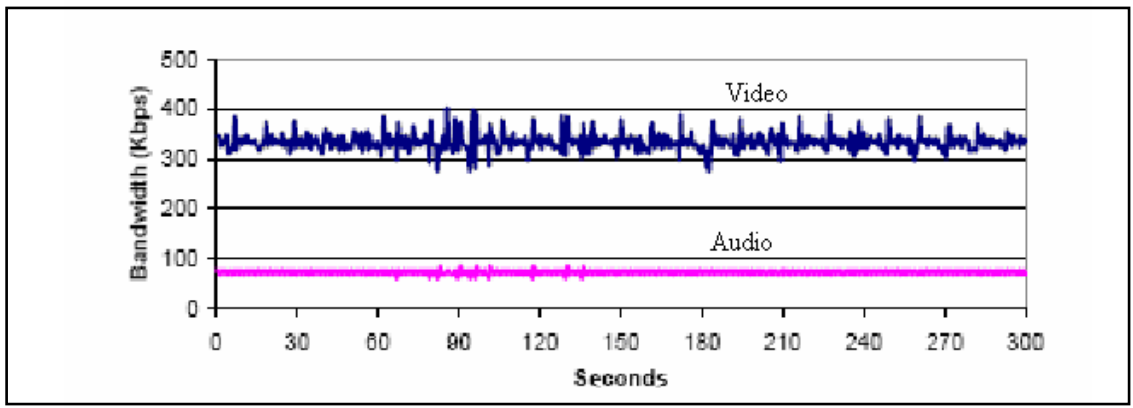

Figure 2.2: Bandwidth utilization during a typical videoconferencing session

In real-time communications the receiver need not have to wait for the whole streams to be downloaded. It starts to play the received stream as soon as some of the packets of audio or video have been received. The media stream is not predefined by duration. It is not necessary to wait for a large durational stream to be downloaded and then play it.

The real-time packets must arrive in time and in general it is not possible for the transport protocol to ask for a lost packet to be resent, and then wait for the source to send it again. The round trip delay to the source and back again can be too long for meeting the time synchronization requirement. So TCP provides no value to these streams, they are carried instead with the User Datagram Protocol (UDP). However UDP has no recovery mechanism. It does not give any guarantee whether the packet will reach or not or the order of their receiving. So it must be ensured that the packets associated with voice or video conferencing arrive the destination in a timely manner, without getting lost, and with no help from the transport protocol. This is the constraint faced by real-time applications.

UDP is a transport layer protocol. On top of this, a lower level protocol which is more application specific has been suggested by AVT working group of IETF, called Real time Transmission Protocol (RTP) [Figure 2.3]. Real time transmission protocol is defined in IETF RFC 1889 [19].

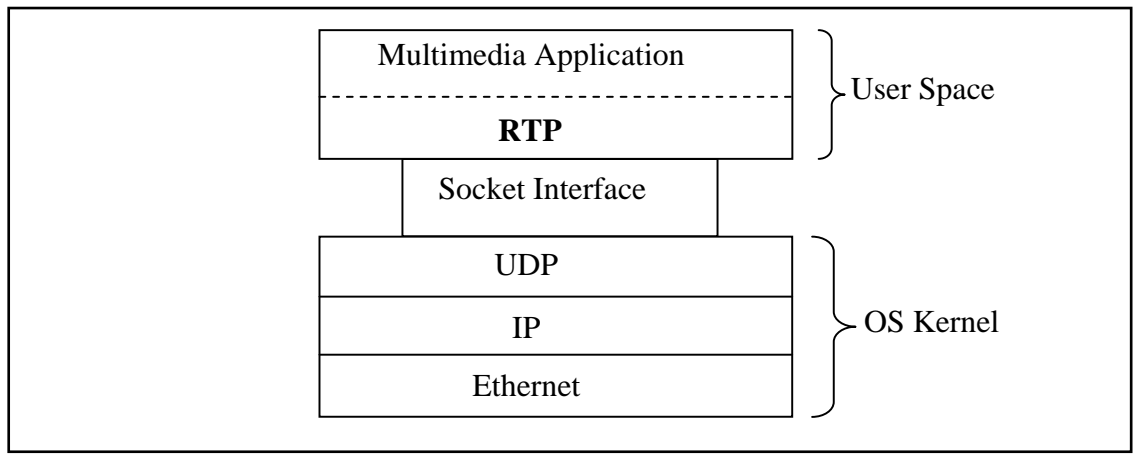

Figure 2.3: The position of RTP in the protocol stack

## 2.2 Real-Time Transmission Protocol (RTP)

RTP provides end-to-end service for transmitting data in real-time. It can be used for both unicast and multicast services. For unicast services separate copy of data are sent from source to destination. For multicast service a source sends only one copy of data. It is the responsibility of the network for transmitting data to multiple locations.

RTP enables to identify the type of data being transmitted, what the order of the packet, synchronies media streams. It dose not give any guarantee of the arrival of the packets. It is the responsibility of receiver to reconstruct and detect loss packets from the information provided by packet header.

It does not provide any mechanism for timely delivery or other QoS guarantee. Real time Transmission Control Protocol (RTCP) enables to monitor quality of data distribution and also provide control and identification mechanism for RTP transmission.

Usually if quality of service is essential for a particular application, RTP is used over RSVP that provides connection oriented service. This is a part of Integrated Service model of QoS implementation.

### 2.2.1 RTP architecture

RTP session is an association of a set of application defined by a network address and a pair of ports (one for media data and other for control data). Each media (voice/video) is transmitted in different sessions. The receiver may choose which session it wants to receive. For example, one receiver has low bandwidth so it will receive only audio session.

**Data packets**

The media is transmitted as a series of packets. A series of packets originating from some source is called the stream. Each packet has two parts – header and actual data.

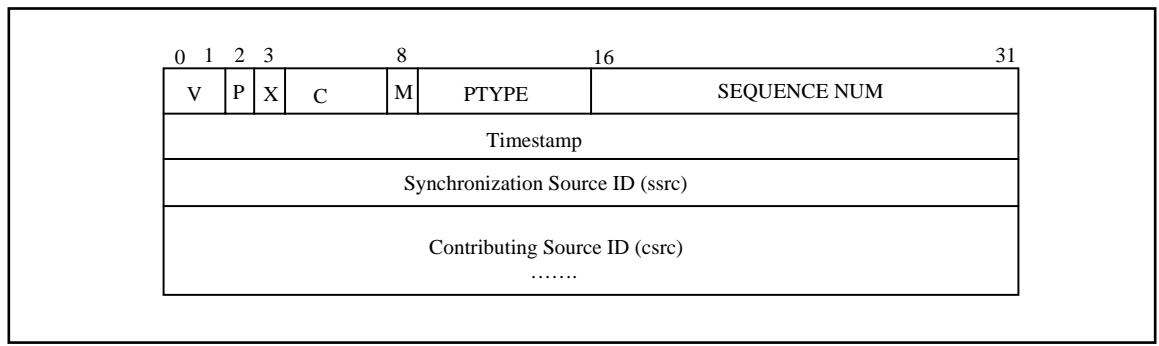

Figure 2.4: RTP packet header

Figure 2.4 shows different fields of a RTP packet header. The header of an RTP data packet contains:

RTP version number (V): 2 bits. The version defined by the current specification is 2.

Padding (P): 1 bit. If the padding bit is set, there are one or more bytes at the end of the packet that are not part of the payload. The very last byte in the packet indicates the number of bytes of padding. The padding is used by some encryption algorithms.

Extension (X): 1 bit. If the extension bit is set, the fixed header is followed by one header extension. This extension mechanism enables implementations to add information to the RTP Header.

CSRC Count (CC): 4 bits. It is the number of CSRC identifiers that follow the fixed header. If the CSRC count is zero, the synchronization source is the source of the payload.

Marker (M): 1 bit. It is a marker bit defined by the particular media profile.

Payload Type (PT): 7 bits. It is an index into a media profile table that describes the payload format. The payload mappings for audio and video are specified in RFC 1890 [20].

Sequence Number: 16 bits. This is a unique packet number that identifies this packet's position in the sequence of packets. The packet number is incremented by one for each packet sent.

Timestamp: 32 bits. It reflects the sampling instant of the first byte in the payload. Several consecutive packets can have the same timestamp if they are logically generated at the same time, for example, if they are all part of the same video frame.

Synchronization Source ID (SSRC): 32 bits. It identifies the synchronization source. If the CSRC count is zero, the payload source is the synchronization source. If the CSRC count is nonzero, the SSRC identifies the mixer.

Contributing Source ID (CSRC): 32 bits each. It identifies the contributing sources for the payload. The number of contributing sources is indicated by the CSRC count field; there can be up to 16 contributing sources. If there are multiple contributing sources, the payload is the mixed data from those sources.

**Control packets**

In addition to data packets RTCP packets are sent periodically. A RTCP packet contains the information about QoS for the session and source and statistics of data that has been transmitted. RTCP packets are stackable and are sent as a compound packet that contains at least two packets, a report packet and a source description packet. All participants in a session send RTCP packets. There are several types of RTCP packets:

Sender Report: A participant that has recently sent data packets issues a sender report. The sender report (SR) contains the total number of packets and bytes sent as well as information that can be used to synchronize media streams from different sessions.

Receiver Report: Session participants periodically issue receiver reports for all of the sources from which they are receiving data packets. A receiver report (RR) contains information about the number of packets lost, the highest sequence number received, and a timestamp that can be used to estimate the roundtrip delay between a sender and the receiver.

The first packet in a compound RTCP packet has to be a report packet. If no data has been sent or received in which case, an empty receiver report is sent.

Source Description: It contains the canonical name (CNAME) that identifies the source. Additional information might be included in the Source Description, such as the source's name, email address, phone number, geographic location, application name, or a message describing the current state of the source.

Bye: When a source is no longer active, it sends an RTCP BYE packet. The BYE notice can include the reason that the source is leaving the session.

Application-specific: RTCP APP packets provide a mechanism for applications to define and send custom information via the RTP control port.

## 2.2.2 RTP applications

RTP applications are divided into RTP Clients - those are able to receive data from the network and RTP Servers - those are able to transmit data across the network. Some applications do both. For example, conferencing applications capture and transmit data at the same time that they are receiving data from the network.

**Receiving media streams from the network**
Being able to receive RTP streams is necessary for several types of applications.
For example:
-Conferencing applications need to be able to receive a media stream from an RTP session and render it on the console.
-A telephone answering machine application needs to be able to receive a media stream from an RTP session and store it in a file.
-An application that records a conversation or conference must be able to receive a media stream from an RTP session and render it on the console and store it in a file.

**Transmitting media streams across the network**
RTP server applications transmit captured or stored media streams across the network. For example, in a conferencing application, a media stream might be captured from a video camera and sent out on one or more RTP sessions. The media streams might be encoded in multiple media formats and sent out on several RTP sessions for conferencing with heterogeneous receivers. Multiparty conferencing could be implemented without IP multicast by using multiple unicast RTP sessions.

## 2.3 QoS in different types of real-time traffic

It has been already stated that real-time transmission is more delay and loss sensitive than non real-time traffic. Some examples of real-time traffic are video conferencing, voice over IP (VoIP), remote class room, tele-haptic, internet gaming etc. In real-time traffic several video and voice codec are used such as IEEE standard MPEG4, MPEG2 , ITU standard based H264, H263, H261 for video and G711, G723, GSM for voice transmission. The video transmission is of variable bit rate (VBR-rt) and variable packet size. Voice transmission is of constant data rate (CBR-rt) and fixed packet size. There are many factors for packet delay and packet loss for real-time transmission. The most important ones are delay and packet loss caused by buffering at the network device such as switches and routers. The distortion of voice and video can be caused by packet loss and delay. So the proper QoS assurance should be there for providing good quality real-time interactive transmission. Conventional IP networks treat all the traffic as Best Effort. And it does not guarantee actual and timely delivery. There are different mechanisms for Quality of Service implementation for real-time traffics which will be discussed in the next chapter.

# Chapter 3

# Quality of Service in Real-Time Communication

In this chapter the Quality of Service model for packet data networks has been discussed. First the description of QoS is presented and how it can be implemented in real-time communication, is described. Then different approaches of QoS in IP networks are discussed and finally a QoS mechanism as deployed in this thesis work is presented.

## 3.1 What is Quality of Service (QoS)?

Quality of Service (QoS) implies the statistical performance guarantee of a network system. It may be defined by some parameters such as average packet loss, average delay, average jitter (delay variation) and average throughput. QoS is the mechanism deployed in real-time transmission to give same sort of priority to the voice and video streams, to insure they will be delivered correctly. The term QoS has been used in many different ways of providing better service to some types of traffic in a network environment, such as priority queuing, application specific routing, bandwidth management, traffic shaping and many others. The QoS implementation may be divided into two broad categories:

1) Application layer QoS – These are implemented at the application level i.e. at the end system. The jitter is mainly controlled at end system in such cases.

2) Network layer QoS – In this approach the QoS performance are controlled at the network layer i.e. at the router and switch. Usually the bandwidth compliance and delay is controlled in the intermediate system in such cases.

In this thesis work network level QoS has been considered and different types of QoS mechanism are discussed in the following sections.

## 3.2 QoS implementation at IP networks

As mentioned earlier the network must have a QoS mechanism for real-time traffic that operates at each switch and router to prioritize real-time traffic. A number of different mechanisms exist in modern networks, such as IntServ (RSVP), DiffServ, IEEE 802.1p/q, and IP Precedence. To provide the maximum benefit to any application, QoS must work from end-to-end. All routers and switches between the real-time sender and the real-time receiver must have a QoS mechanism available and enabled.

QoS is implemented both at layer 2 and 3 in the protocol stack. Layer 3 QoS will be recognized and handled by routers in the network. Since congestion also occurs in switch output queues, so level 2 QoS is also required.

There are two methodologies available in most networks for implementing QoS at layer 3
- Integrated Service Model (IntServ)
- Differentiated Service Model(DiffServ)

### 3.2.1 Integrated Service (IntServ) model

This model assumes that each node in the network is QoS aware. It provide support for two broad class of services, real-time applications with strict bandwidth and latency requirements can be processed by guaranteed service while the traditional applications which require performance equivalent to lightly loaded network are provided controlled load service [1]. To negotiate the QoS with an IntServ capable network signaling and admission control is used.

The main Internet Engineering Task Force (IETF) protocol for Integrated Service architecture is RSVP (Resource reSerVation Protocol) described in RFC 2205[2]. The host use RSVP protocol to request specific QoS from the network for particular application. It is also used by the routers to deliver QoS request to all nodes along the paths of flows. RSVP requests will generally result in resources being reserved in each node along the data path. The main features of RSVP are as follows:

- RSVP makes resource reservations for both unicast and multicast applications, adapting dynamically to changing group membership as well as to changing routes.
- It makes reservations for unidirectional data flows.
- The receiver of a data flow initiates and maintains the resource reservation used for that flow.
- The reserved resources are – bandwidth, buffer space(router buffer), CPU cycles (router CPU time to process the packets)
- RSVP requires each router to maintain information about each active real-time stream. So this solution is not scalable for large networks.

IntServ has problems of scalability and is not yet widely accepted in deployment across the internet. In this project currently Differentiated Service model has been considered for implementation.

## 3.2.2 Differentiated Service (DiffServ) model

Differentiated Service (DiffServ) as defined in RFC-2475 [3] was developed as an alternative approach that would have better scaling properties. Rather than specifying resources for each real-time stream, DiffServ allocates resources for a class of traffic. All traffic allocated to that class is treated with the same policy, such as being queued in a high priority queue.

DiffServ is provided by the router and defines a set of service classes with corresponding forwarding rules. A packet coming to the router with a Type of Service (ToS) field may get better services than other, provided by some classes depending upon that ToS content. There are different forwarding classes as follows.

a) Expedited Forwarding (EF) – It is defined in RFC-3246 [22]. It is intended to provide a building block for low delay, low jitter and low loss services by ensuring that the traffic is served at a certain configured rate over a suitably defined interval, independent of the offered load of non-EF traffic to that interface.

b) Assured Forwarding (AF) – It is defined in RFC-2597 [23]. The AF Per Hop Behavior (PHB) group provides delivery of IP packets in four independently forwarded AF classes. Within each AF class, an IP packet can be assigned one of three different levels of drop precedence, low, medium and high. So 12 classes of services may be possible in AF as in Table 3.1.

Table 3.1: DSCP for different class of services

|  | Class 1 | Class 2 | Class 3 | Class 4 |
|---|---|---|---|---|
| **Low Drop Prec** | 001010 | 010010 | 011010 | 100010 |
| **Medium Drop Prec** | 001100 | 010100 | 011100 | 100100 |
| **High Drop Prec** | 001110 | 010110 | 011110 | 100110 |

Four AF classes are defined, where each AF class is in each node allocated a certain amount of forwarding resources (buffer space and bandwidth).Within each AF class IP packets are marked with one of three possible drop precedence values. In case of congestion, the drop precedence of a packet determines the relative importance of the packet within the AF class. A congested node tries to protect packets with a lower drop   precedence value from being lost by preferably discarding packets with a higher drop precedence value. In a node, the level of forwarding assurance of an IP packet thus depends on
(1) How much forwarding resources has been allocated to the AF class that the packet belongs to?
(2) What is the current load of the AF class in case of congestion within the class?
(3) What is the drop precedence of the packet?

Expedited Forwarding (EF) is recommended for voice transmission, which is implemented with a priority queue. The priority queue is always emptied before other queues, giving the lowest possible latency to that traffic. The video conferencing is typically implemented using a high priority Assured Forwarding (AF) marking, which is implemented with a rate-based queue. This is appropriate for a converged network where both voice and video conferencing are being implemented, because it ensures that the voice traffic has priority over the higher bandwidth and bigger packets of the video conferencing stream.

### 3.2.3 Differentiated Services Code Point (DSCP)

The most significant 6 bits in the type of service field of the IP header are used by Differentiated Services and known as Differentiated Services Code Point (DSCP). $64(2^6)$ different DSCP values may be possible. Figure 3.1 shows the position of ToS and DSCP field in the IP header. Table 3.2 is the list of DSCP value for different forwarding classes.

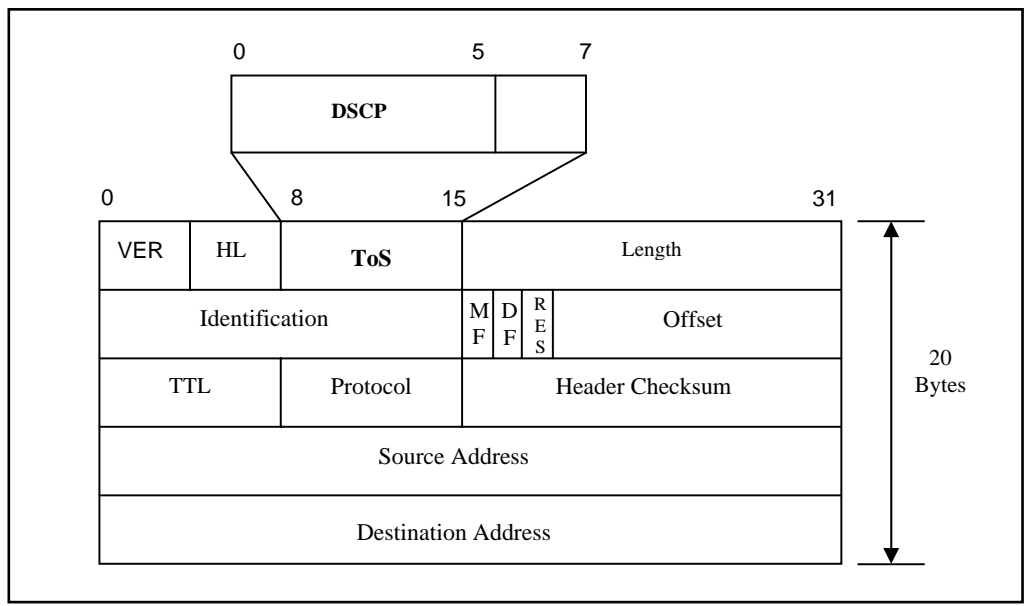

Figure 3.1: Type of Service (ToS) field in IP header

Table 3.2: DSCP values for different forwarding classes

| DSCP | | Forwarding classes |
|---|---|---|
| 0 | 000000 | Best Effort |
| 8 | 001000 | Class 1 |
| | 001010 | AF11 |
| | 001100 | AF12 |
| | 001110 | AF13 |
| 16 | 010000 | Class 2 |
| | 010010 | AF21 |
| | 010100 | AF22 |
| | 010110 | AF23 |
| 24 | 011000 | Class 3 |
| | 011010 | AF31 |
| | 011100 | AF32 |
| | 011110 | AF33 |
| 32 | 100000 | Class 4 |
| | 100010 | AF41 |
| | 100100 | AF42 |
| | 100110 | AF43 |
| 46 | 101110 | Expedited Forwarding |
| 48 | 110000 | Network Control |

IP Precedence is a methodology created in the original IP specification. It is a much simpler mechanism that gives precedence to IP packets marked as high priority. The bit positions in the IP header formerly used for IP Precedence and Type of Service (ToS) have been reassigned for use with DiffServ [24].

The technology in use for providing QoS at layer 2 is IEEE 802.1p. IEEE 802.1p functionality is often coupled with IEEE 802.1q, which is the VLAN function. Both technologies use the same bit field added to the Ethernet header, to specify a streams priority and VLAN association. Switches implementing IEEE 802.1p use multiple output queues for each switch output port. Higher priority traffic is assigned to a higher priority queue, and those queues are serviced before lower priority queues.

### 3.2.4 Asynchronous Transfer Mode (ATM)

Asynchronous Transfer Mode (ATM) provides a network layer QoS mechanism and was created by Asynchronous Transfer Mode Forum [17]. The most widely available ATM service categories are

- Constant Bit Rate (CBR) – used for circuit emulation services including circuit based voice and video transport.
- Real-time Variable Bit Rate (VBR - rt) – used for real-time packet based voice or video services.
- Non real-time Variable Bit Rate (VBR - nrt) – used for priority data service.
- Unspecified Bit Rate (UBR) – used for best effort data service.

Table 3.3 gives the mapping of QoS from DiffServ domain to ATM service.

Table 3.3: DiffServ to ATM QoS mapping

| **DiffServ Code Point (DSCP)** | **ATM Service Category** |
|---|---|
| CS7, CS6, CS5, EF | CBR or VBR – rt |
| AF4x, CS4, AF3x, CS3 | VBR – rt |
| AF2x, CS2, AF1x, CS1 | VBR – nrt |
| DE, CS0 | UBR |

ATM service categories are not the same as DiffServ categories, but the DiffServ categories can be mapped into ATM categories at the network boundary. Table – 4 shows the mapping between DiffServ code points and ATM service categories.

## 3.2.5 Multi Protocol Label Switching (MPLS)

MPLS is an IETF–specified framework that provides for the efficient designation, routing, forwarding, and switching of traffic flows through the network [17]. It is placed between layer 2 and layer 3 as shown in Figure 3.2.

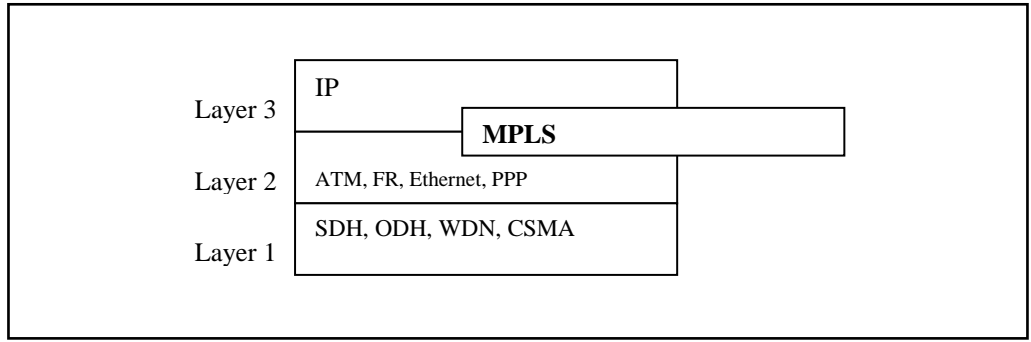

Figure 3.2: MPLS layer in between layers 2 and 3

MPLS performs the following functions:

- Specifies mechanisms to manage traffic flows of various granularities, such as flows between different hardware, machines, or even flows between different applications.
- Remains independent of the layer-2 and layer-3 protocols.
- Provides a means to map IP addresses to simple, fixed-length labels used by different packet-forwarding and packet-switching technologies.
- Interfaces to existing routing protocols such as Resource reSerVation Protocol (RSVP) and Open Shortest Path First (OSPF).
- Supports the IP, ATM, and frame-relay layer-2 protocols

## 3.3 How to setup QoS in IP networks for real-time communication

For real-time traffic differentiated service based QoS mechanism is most scalable approach. This mechanism is deployed in two steps:
  a) The real-time packets i.e. voice and video packets will be marked by different DSCP value either by the system where is has been generated or at the switch through which it will go to the router where QoS will be implemented on these packets. In this work currently the marking of packets is assumed to be done at the switch level.

b) Depending upon this DSCP value the packets are distributed among different class of service and different scheduling policy will be implemented on those classes to achieve a desired QoS.

Figure 3.3 describes different elements in a modern router to implement QoS.

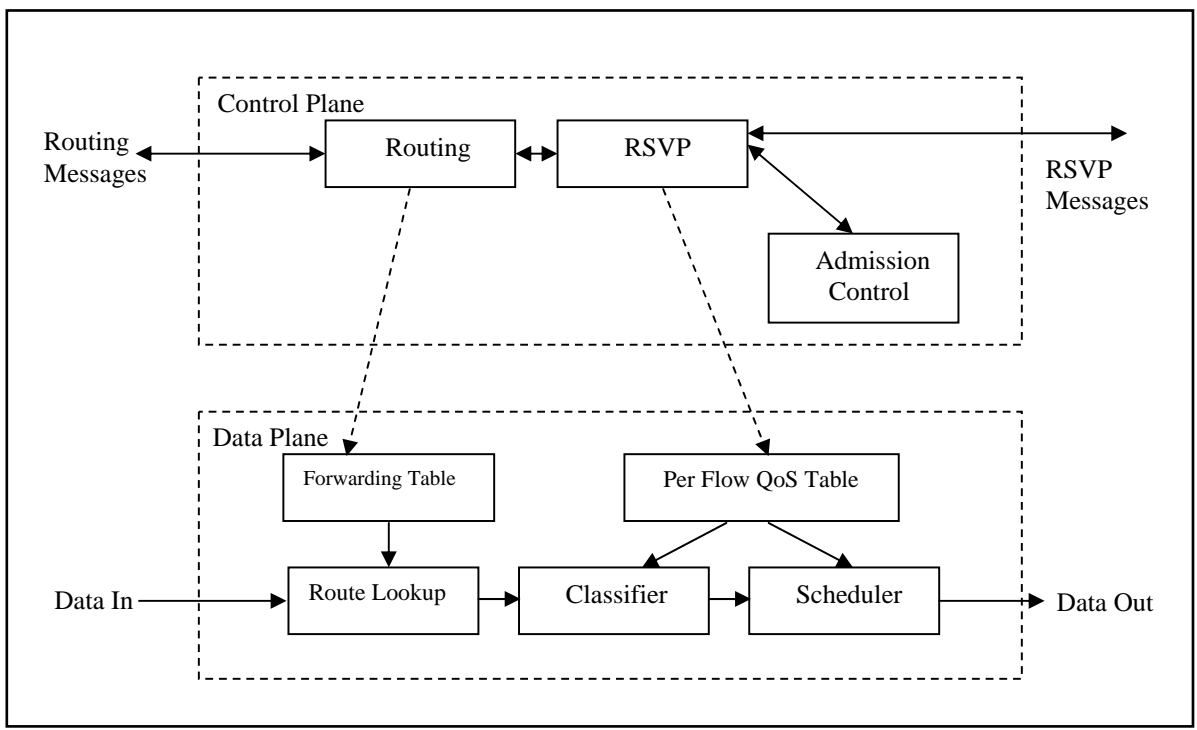

Figure 3.3: Different elements in modern router

A working mechanism of router may be viewed as different planes, like

- Control Plane
- Data plane

In Figure 3.3 control plane and data plane has been depicted. Resource reservation request and admission control are done in control plane. Data plane is responsible for packets classification and scheduling.

Packets are classified depending upon the DSCP value of packet header and assigned to different queues (buffers) of forwarding classes as shown in Figure 3.4.

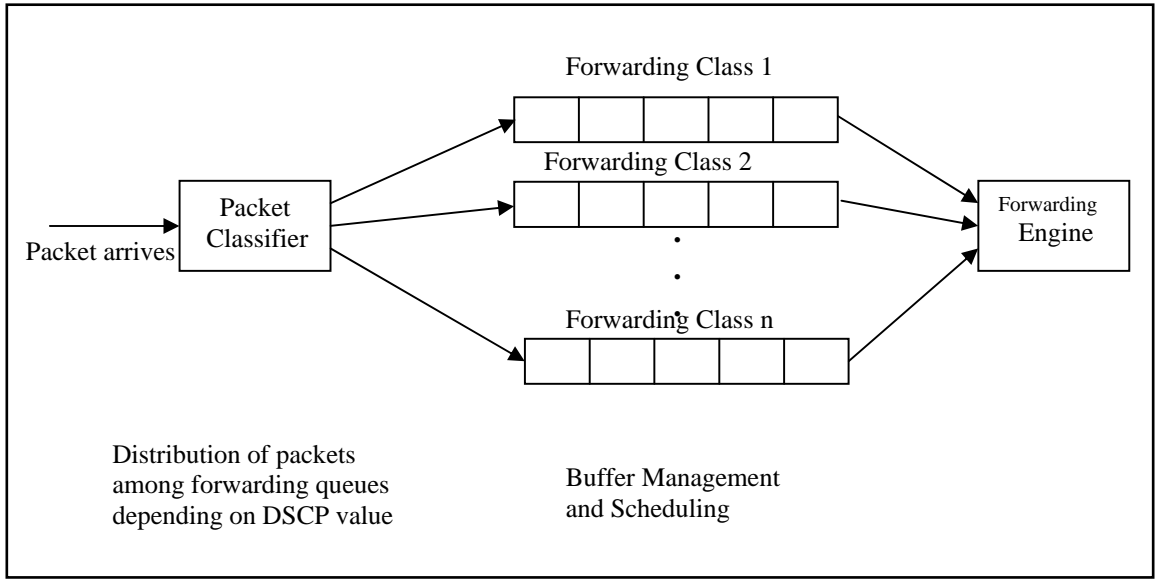

Figure 3.4: Classification of packets

Queues are the primary contributors to packet loss and delay in a packet data network. There are queues at each output port in each router and switch in the network, and packets have to enter and leave an output queue on each device through which it passes on its route. If queues are empty or nearly empty, the packet enters and is quickly forwarded onto the output link. If momentarily traffic is heavy, queues fill up and packets are delayed waiting for all earlier packets in the queue to be forwarded before it can be sent to the output link. If the traffic is too high, the queue fills, and then packets are discarded, or lost.

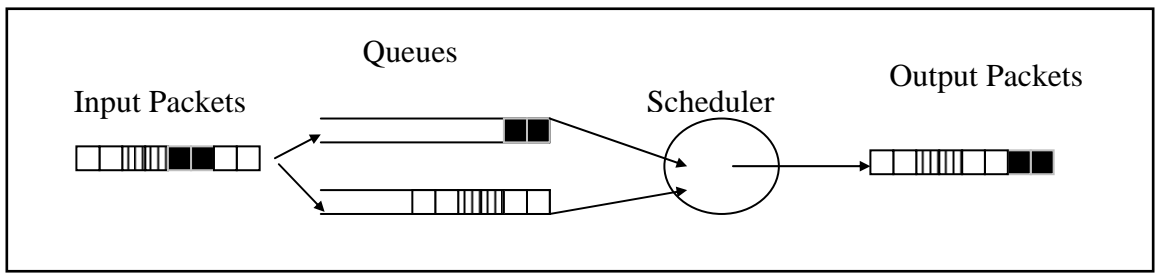

Figure 3.5: Packet queue diagram

A priority queuing mechanism provides additional queues at each switch or router output port, dedicated to high priority traffic. In Figure 3.5 a simple 2-queue output port is depicted. Best effort traffic are queued in the lower queue, while high priority traffic, here colored black, is queued in the upper queue. The queue policy determines how the queues will be emptied when they both have packets waiting. A high priority queue is always emptied before any lower priority queue. So, in Figure 3.5 both black packets are serviced first, and then the remaining packets in the best effort queue. If additional high-priority packets arrive while the lower queue is being emptied, service is immediately switched to the high priority queue.

There are several packet scheduling schemes, such as,
- First Come First Serve (FCFS) – The packets arrives first in the queue is forwarded first.
- Priority Scheduling (PS) - The queues are assigned with some priority value. And the high priority queue is emptied before the lower priority one.
- Weighted Round Robin (WRR) – It empties queues based on the bandwidth that has been allocated to each queue. If queue 1 is allocated 40% of the available bandwidth, and queue 2 is allocated 60% of the available bandwidth, then queue 2 is serviced 6/4 times as often as queue 1.

    There are different ways of buffer management mechanism, such as
- Push out method – high priority packet which arrives to a full buffer will discard the low priority one.
- Threshold method – discard low priority packets if the buffer content is above a certain level.

There is a tradeoff between queue length and packet drop. If the queue length is small there may be packet drop due to burst traffic though the waiting time for the small buffer will be low. On the other hand for large buffer there will be no packet drop but the waiting time will be high. Random Early Detection (RED) algorithm is used to get better service in both the above cases. RED determines that the queue length is getting long, and then randomly selects packets within the queue for discard. So there is no need to wait for queue overflow or packet drop in burst.

This thesis work concentrates in developing a DiffServ based QoS model for real-time interactive traffic. The model is implemented at the edge of the network. It assumes the availability of the requisite form of routers and an idea of different types of traffic originating or receiving by end users of the network. The suggested QoS model and the testing results will be presented from the next chapter onwards.

# Chapter 4

# A Traffic Engineering Model for QoS Provisioning in Networks at Edge Routers

This chapter describes a traffic engineering model that has been proposed to deploy the QoS for real-time interactive traffic in a network. The model is based on the DiffServ based QoS as mentioned in the chapter 3 and has been implemented on the routers at the sender edge of the network. The bandwidth at the edge of the network is shared by many users and can be constrained by traffic volume.

The objective of the thesis work is to develop and deploy a traffic engineering model to provide good QoS to real-time interactive traffic with the constraint of available resources. Some well known QoS mechanisms for real-time traffic are Resource reSerVation Protocol (RSVP), Integrated Services (IntServ), Differentiated Services (DiffServ), and Multi Level Switch Protocol (MPLS). These are already described in the previous chapter. Among these the most scalable QoS mechanism is DiffServ which has been chosen for providing good QoS to the real-time interactive traffic. The DiffServ has been implemented at the edge of the network of sender side. The proposed model has been simulated for verification. The description of this model and simulation results are presented in the following sections.

## 4.1 Proposed traffic engineering model

The aim of the proposed QoS model is to reduce the packet loss and packet delay for real-time interactive traffic with the constraint of available resources. The real-time traffic may be of Constant Bit Rate (CBR-rt) and Variable Bit Rate (VBR-rt). The video portion of a video conferencing session may be considered as VBR-rt and the voice portion of Voice over IP (VoIP) traffic may be considered as the CBR-rt traffic for the proposed model. The real-time traffic is generated at the sender side and is sent to the destination through QoS configured network device at the edge of the network. The network devices are consisting of switch and routers.

DiffServ QoS mechanism has been considered for aggregated class of traffic. The proposed QoS mechanism has been deployed at the edge switch and edge router of the IP network with bottleneck link at the sender side. Most of the routers provide 4 logical queues to support DiffServ QoS mechanism. One queue is considered as a high priority queue, which is used for some signalling traffic. This queue is called network control queue. The proposed QoS model assumes the existence of this queue apart from the other queues and implements QoS with only the remaining 3 queues. Figure 4.1 is the block diagram of the QoS mechanism that can be used for the egress traffic at the edge of the network.

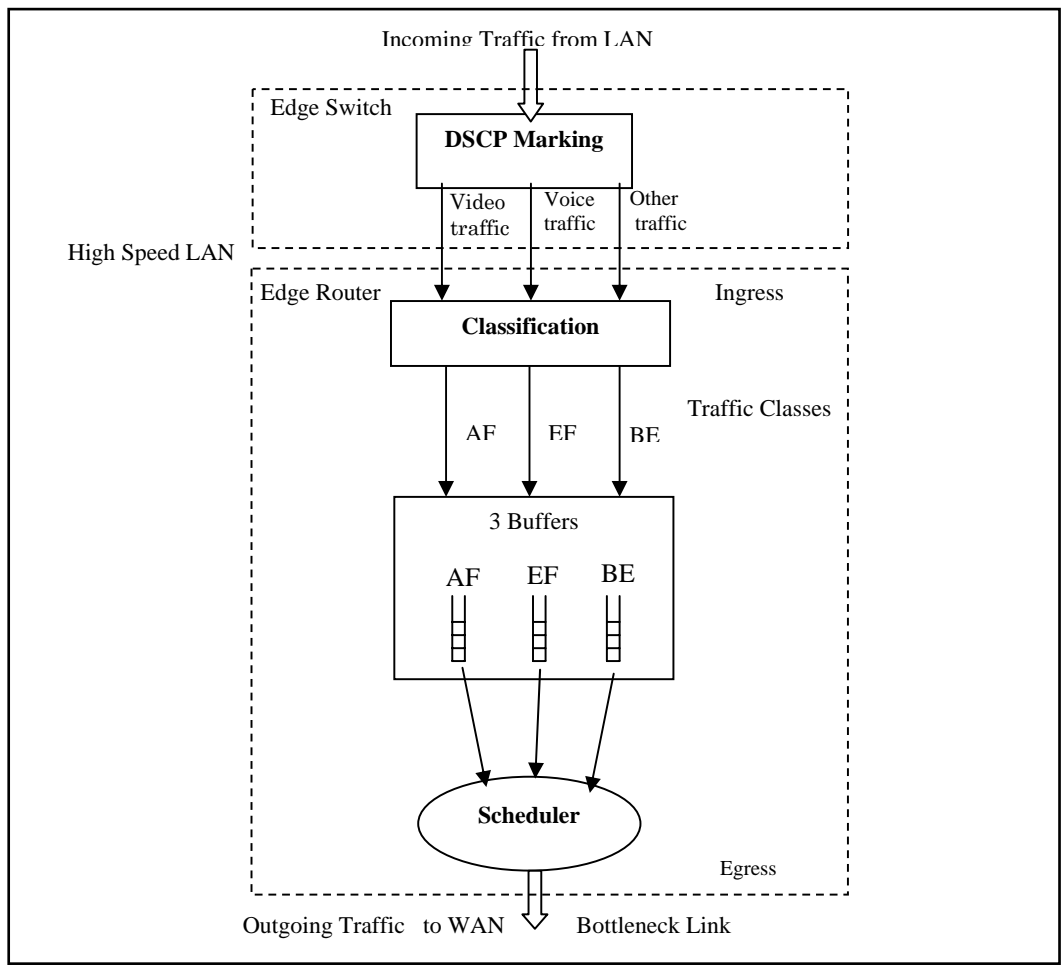

Figure 4.1: Scheme for implementing QoS on egress traffic at the edge of a network

The DiffServ mechanism can be divided into following three steps –

(i) Marking: The incoming traffic from LAN are marked by different DSCP value for different types of traffic, e.g., voice traffic are marked by Expedited Forwarding(EF) DSCP value, video traffic are marked by Assured Forwarding(AF) DSCP value and rest of the non real-time traffic are marked as default Best Effort (BE) DSCP value. Marking is done at the edge routers of the network at the sender side.

(ii) Classifying: Marked traffic are classified among four classes depending upon the DSCP value.

Assured Forwarding (AF) class: Video traffic marked with AF DSCP value goes to this class.

Expedited Forwarding (EF) class: Voice traffic marked with EF DSCP value goes to this class.

Best Effort (BE) class: Rest of the traffic (non real-time) goes to this class.

Classifying is done at core routers of the network at the sender side.

(iii) Scheduling: The routers assign logical queues to each of the traffic classes and the available bandwidth is distributed among the queues with some scheduling mechanism. Queues are the primary contributors to packet loss and delay in a packet network. In this thesis work a QoS scheduling model has been designed and validated with simulation as well as experimental results. The model is described below.

### QoS scheduling model

In this QoS scheduling model the total available buffer and bandwidth are distributed to different classes of services. The model is dynamic in nature as the buffer allocation procedure is adaptive with current queue length i.e., current offered load to that queue. Bandwidth allocation procedure uses buffer size as weightage of Weighted Round Robin (WRR) mechanism. Service rate of the queue is proportional to the allocated bandwidth to that queue. In this model first the QoS scheduling parameters are defined then the allocated resources are derived for each queue. The QoS scheduling model and the supporting parameters are depicted in Figure 4.2.

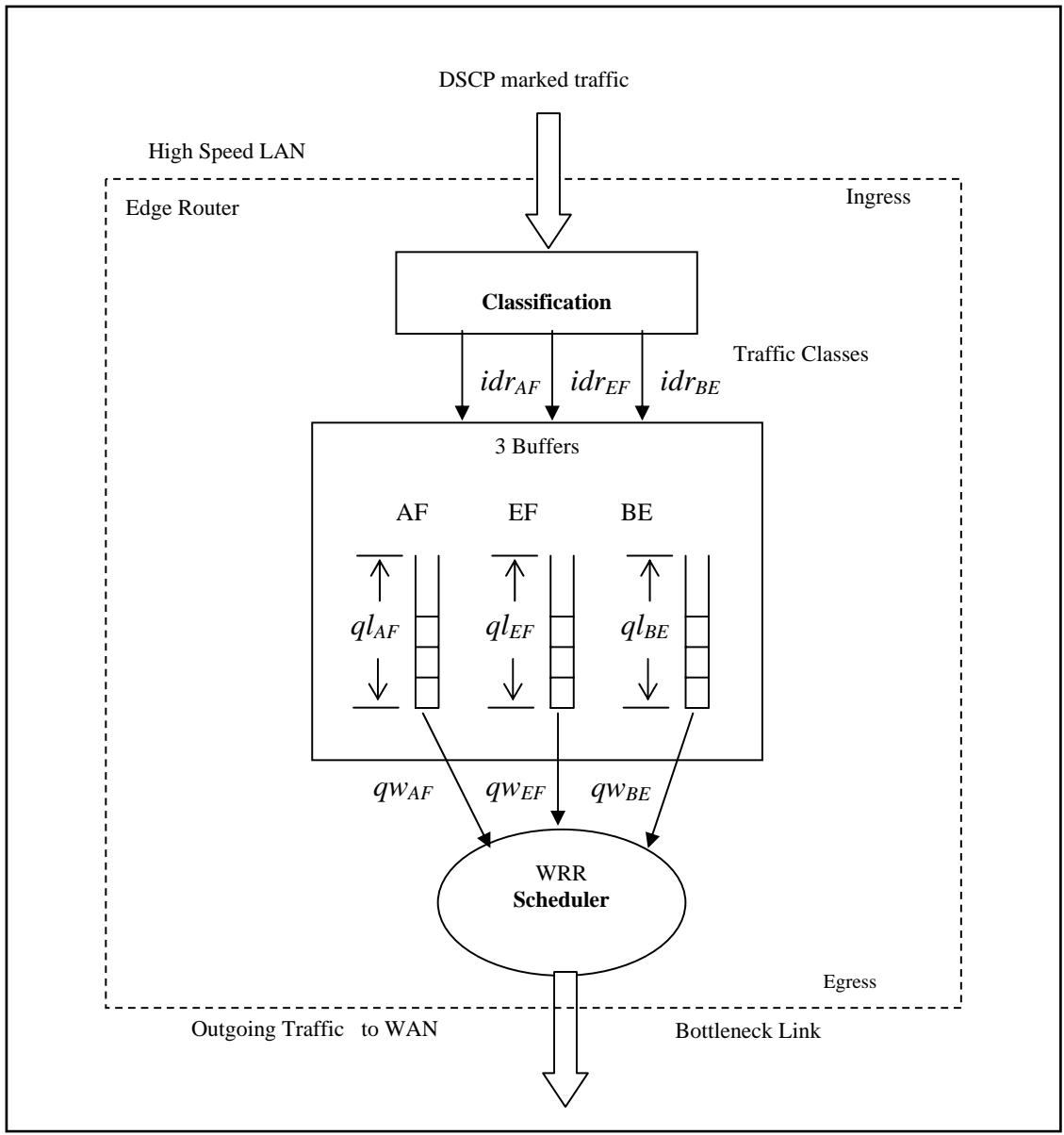

Figure 4.2: QoS Scheduling model with supporting parameters

(A) QoS scheduling parameter:

There are two sets of parameters which control the packet delay and packet loss at the queue of the outgoing traffic. These are,

1) Queue (buffer) length:

$ql_{AF}$ : average queue length or buffer size for AF traffic.

$ql_{EF}$ : average queue length or buffer size for EF traffic.

$ql_{BE}$ : average queue length or buffer size for BE traffic.

2) Queue weight for Weighted Round Robin (WRR) scheduler:

$qw_{AF}$: queue weightage or service rate for AF traffic.

$qw_{EF}$: queue weightage or service rate for EF traffic.

$qw_{BE}$: queue weightage or service rate for BE traffic.

These parameter sets are computed in terms of the following variables.

$idr_{AF}$ = average input data rate for AF traffic.

$idr_{EF}$ = average input data rate for EF traffic.

$pktsz_{AF}$ = maximum packet size for AF traffic.

$pktsz_{EF}$ = average packet size for EF traffic.

$dly_{AF}$ = allowable packet delay for AF traffic.

$dly_{EF}$ = allowable packet delay for EF traffic.

$N_{AF}$ = number of sessions for AF traffic.

$N_{EF}$ = number of sessions for EF traffic.

$BUFF$ = total available buffer at router.

$TB$ = total available bandwidth at the edge network.

Input data rates of AF and EF traffic are collected by observing the characteristics of real-time traffic. Allowable packet delay has been obtained from the QoS requirements of the real-time traffic. Numbers of sessions are given by the admission control mechanism which is assumed to be present before the proposed DiffServ QoS model. Number of session is a function of total available bandwidth at the edge network ($TB$), total available buffer at the router ($BUFF$), average input data rate of real-time traffic ($idr_{AF}$ and $idr_{EF}$) and average packet size of the real-time traffic ($pktsz_{AF}$ and $pktsz_{EF}$). Total available buffer at the router are found from the router specification.

(B) Buffer allocation for each class of traffic:

The allocated buffers for different queues are computed as follows,

$ql_{AF} = (idr_{AF} * dly_{AF} * N_{AF}) / pktsz_{AF.}$  (1)

$ql_{EF} = (idr_{EF} * dly_{EF} * N_{EF}) / pktsz_{EF}.$  (2)

$ql_{BE} = BUFF - (ql_{AF} + ql_{EF}).$  (3)

(C) Service rate or weightage for each class of traffic for WRR:

The scheduling policy is based on Weighted Round Robin (WRR) algorithm. The queues are served according to the queue weight of a particular class of traffic. Hence, the queue weight is same as service rate for the corresponding queue. Queue weight is adaptive with the current queue length of particular class of traffic. The time interval considered in determining queue length is assumed sufficiently large to capture some minimum BE traffic. If the queue lengths of three classes are same for a particular time the scheduling algorithm goes with priority of the class of traffic. EF priority is higher than AF priority. The weightage assignment is done in the following way.

Let, $qw_{AF}(t)$ = service rate or queue weightage for AF traffic at time t.

$qw_{EF}(t)$ = service rate or queue weightage for EF traffic at time t.

$qw_{BE}(t)$ = service rate or queue weightage for BE traffic at time t.

$ql_{AF}(t)$ = queue length or buffer size for AF traffic at time t.

$ql_{EF}(t)$ = queue length or buffer size for EF traffic at time t.

$ql_{BE}(t)$ = queue length or buffer size for BE traffic at time t.

$P_{AF}$ = AF traffic priority = 2.

$P_{EF}$ = EF traffic priority = 3.

$P_{BE}$ = BE traffic priority = 1.

Now the computed service rates (using equations 1, 2 and 3) are,

$qw_{AF}(t) = (ql_{AF} * P_{AF}) / (ql_{AF}(t) + ql_{EF}(t) + ql_{BE}(t))$.    (4)

$qw_{EF}(t) = (ql_{EF} * P_{EF}) / (ql_{AF}(t) + ql_{EF}(t) + ql_{BE}(t))$.    (5)

$qw_{BE}(t) = (ql_{BE} * P_{BE}) / (ql_{AF}(t) + ql_{EF}(t) + ql_{BE}(t))$.    (6)

Equations 4, 5 and 6 show that the service rate of the queue is adaptive to the ratio of length of the particular service queue to the total length of the all service queues taken together. If the length of a queue class is longer than other queues, that queue will be served faster than the queue whose length is smaller. If the queues are of equal length, they will be served according to their priority value.

Each traffic class therefore is serviced as per the product of queue length and assigned priority. For example voice traffic in EF class will be serviced fastest when its queue length is large. Video traffic in AF class will be serviced faster only if when its queue length is higher. Whereas BE traffic will serviced highest when there is no AF and EF class traffic.

(D) Procedure for serving the queues:

As stated earlier, the QoS scheduling model uses WRR algorithm to serve the queues. The weightage or service rates calculated in the equations 4, 5 and 6 are used as the weightage factor to this WRR algorithm. The scheduler shown in the Figure 4.2 serves different traffic from different queues, proportional to the weightage factor of the corresponding queues and sends it to the forwarding engine and eventually to the network.

The procedure to implement the scheduler can be summarized as follows:

Step1 : Determine the number of sessions for the real-time traffic ($N_{AF}$ and $N_{EF}$).
  These are given by the admission control mechanism which is assumed to
  be present before the proposed DiffServ QoS model based scheduler.

Step 2: Determine queue length for different classes of traffic ($ql_{AF}$, $ql_{EF}$ and $ql_{BE}$) using
  equations 1, 2 and 3.

Step 3: Calculate the WRR service rate or queue weightage ($qw_{AF}$, $qw_{EF}$ and $qw_{BE}$) using
  equations 4, 5 and 6.

Step 4: Servicing rate of a queue is proportional to its weightage. The queues are served
  using following logic:
  *If*
    the length of a queue is longer than the other queue
  *Then*
    serve that queue faster than the queue whose length is smaller
  *Else If*
    the queues are of equal length
  *Then*
    serve the higher priority queue faster than the queue whose priority is lower

The QoS scheduling model described above, is simulated using standard Network Simulator(NS2). The following sections will describe the simulation of the model in details and will disscuss the simulation results as well.

## 4.2 Simulation of the proposed model

The described QoS model has been simulated with standard NS2 simulator [25] to show how the QoS of real-time traffic is improved in presence of the proposed QoS model. The performance of this QoS model is observed by measuring the packet delay and packet loss of the real-time traffic at the receiver side [26].

### 4.2.1 Network topology for simulation

A network topology for implementing the proposed QoS model has been simulated as depicted in Figure 4.3. At the sender side r1, r2 and r3 are video source (AF), voice source (EF) and data source (BE) respectively. They are connected to the edge switch r4 by 100 Mbps link. r4 is connected to the edge router r5 by 1 Gbps link. r5 is connected to the network by a bottleneck link bandwidth say MAX (in the simulation program MAX = 2.1 Mbps). In the receiver side the connection is similar to the sender side. The edge router r6 is connected to edge switch r7. Then r7 is connected to the video sink (r8), voice sink (r9) and data sink (r10). Packet marking is done at edge switch (r4) at the sender side. Packet scheduling is done at the edge router (r5) at sender side.

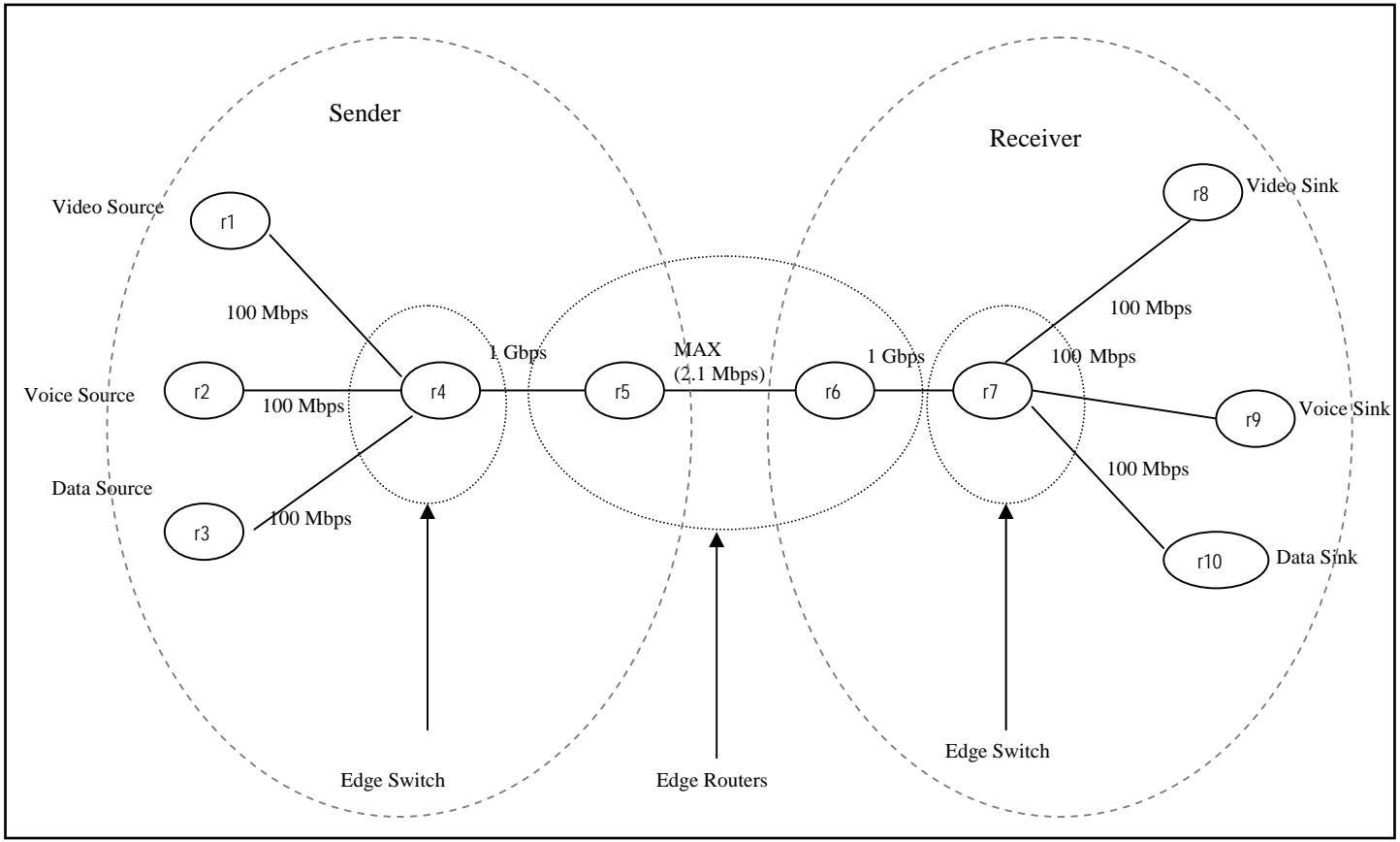

Figure 4.3: Simulated network topology

The video source (r1) is represented as the sequence of data packets containing repeated IBBBPBBB frames generated at predefined fixed interval. The I, B and P frames are however taken to be of variable length following an exponentially distributed function of packet size. It is assumed that the generation of frame sequence is compatible with a fixed time interval of successive I frames. During the simulation time, 3 video sessions are assumed to be running with average data rate 384 Kbps each. The voice source (r2) is represented as CBR traffic source with fixed packet size. 3 voice sessions are assumed to be running throughout simulation time with data rate 64 Kbps each. For data traffic Poisson distributed traffic source (r3) has been considered. The offered traffic load that goes to the bottleneck link of the network is equal to the sum of all aggregated traffic at the queues (aggregated AF traffic load + aggregated EF traffic load + aggregated BE traffic load). The offered load at the link is increased by increasing the Poisson traffic at the BE queue. AF and EF traffic load remain fixed as mentioned above.

## 4.2.2 Simulation results and analysis

QoS performance statistics such as the packet loss and packet delay for each traffic and with different bottleneck links are monitored at the receiver end by increasing the load in the link under the following situations:
1. With enabling proposed QoS model.
2. With disabling proposed QoS model.

Monitoring the performance statistics at the receiver end, it has been seen that the proposed QoS model improves the performance of real-time traffic. Simulation results for various traffic classes are presented in the Figures 4.4 (a, b and c) and Figures 4.5 (a, b and c). The X axis represents the total offered load at the constrained bandwidth bottleneck link. It includes the aggregated AF, EF and BE traffic. The Y axis represents the packet loss [Figures 4.4 (a, b and c)] or packet delay [Figures 4.5 (a, b and c)] of AF, EF and BE traffic.

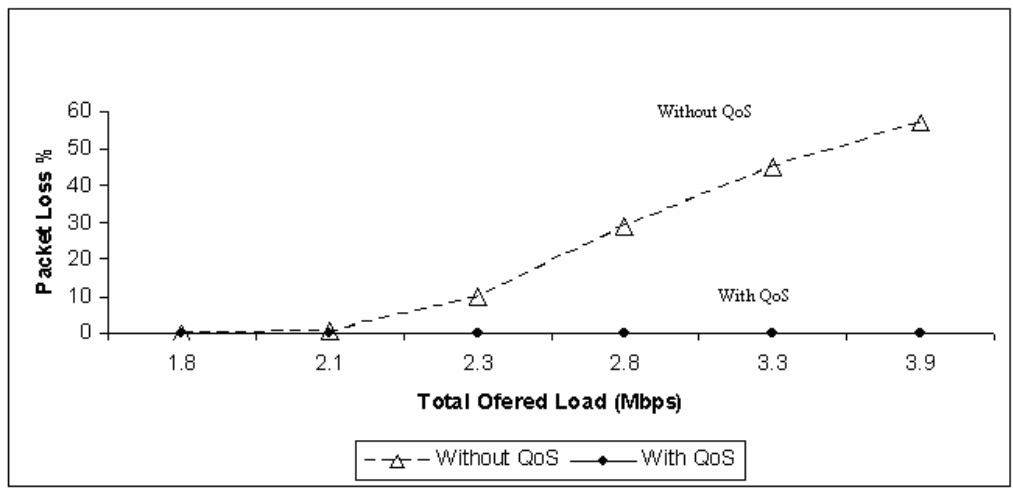

Figure 4.4(a): Packet loss vs. total offered load for AF traffic

Figure 4.4(a) shows that, as the total offered load at the bottleneck link increases, the packet loss at AF queue also increases, when QoS is not enabled in the bottleneck link. On the other hand, enabling QoS at the bottleneck link, the packet loss at AF queue remains at 0%, though the total offered load at the bottleneck link increases beyond the available (constrained) bandwidth at the bottleneck link.

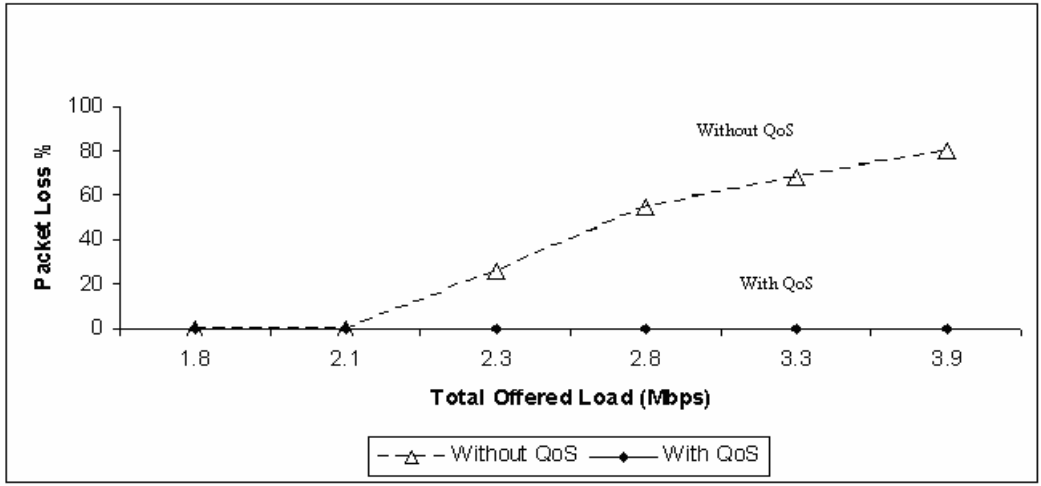

Figure 4.4(b): Packet loss vs. total offered load for aggregated EF traffic

Figure 4.4(b) shows that, as the total offered load at the bottleneck link increases, the packet loss at EF queue also increases, when QoS is not enabled in the bottleneck link. On the other hand, enabling QoS at the bottleneck link, the packet loss at EF queue remains at 0%, though the total offered load at the bottleneck link increases beyond the available (constrained) bandwidth at the bottleneck link.

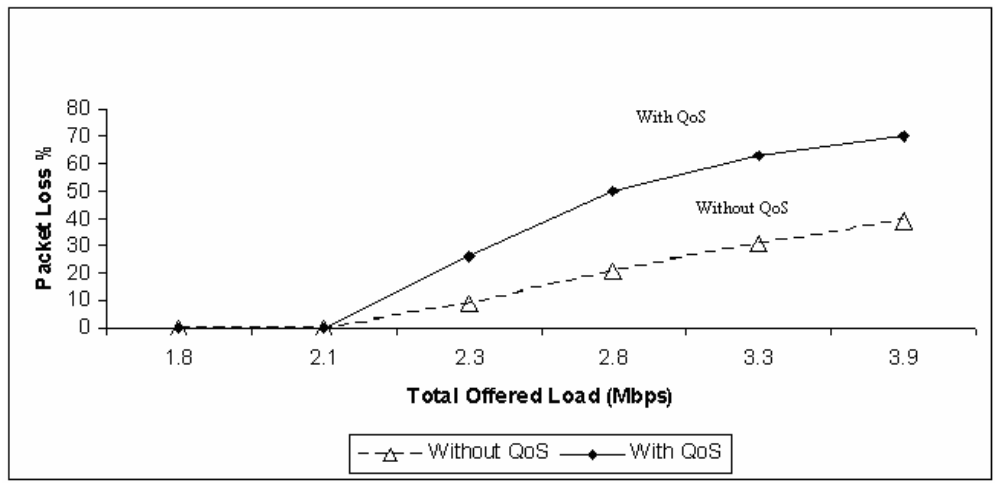

Figure 4.4(c): Packet loss vs. total offered load for BE traffic

Figure 4.4(c) shows that, as the total offered load at the bottleneck link increases, the packet loss at BE queue also increases, when QoS is not enabled in the bottleneck link.

But enabling QoS at the bottleneck link, the packet loss at BE queue increases more rapidly when the total offered load are increasing beyond the available(constrained) bandwidth at the bottleneck link.

It is seen from Figures 4.4 (a, b and c) that the WRR scheduler ensures QoS to limit packet loss for each of the service until the total offered load is up to the bottleneck limit of 2.1 Mbps. Thereafter at higher total offered load the AF and EF packets continue to get guarantee of low packet loss whereas BE traffic gets affected. When total offered load exceeds the bottleneck bandwidth the priorities for AF and EF classes assure higher QoS for the classes repetitively where as the BE traffic class has higher packet loss.

Therefore the QoS mechanism ensures that the real-time traffic has sufficient quality though the total offered load at the bottleneck link is more than available bandwidth. However it is equally important not to disturb BE traffic so that fairness is assured to BE traffic. To achieve this some congestion control mechanism can be implemented for the TCP traffic at the BE queue to provide fairness to it. The discussion for congestion control at BE traffic is not in the scope of this thesis.

Now, the effect of QoS implementation on delay of different traffic will be presented.

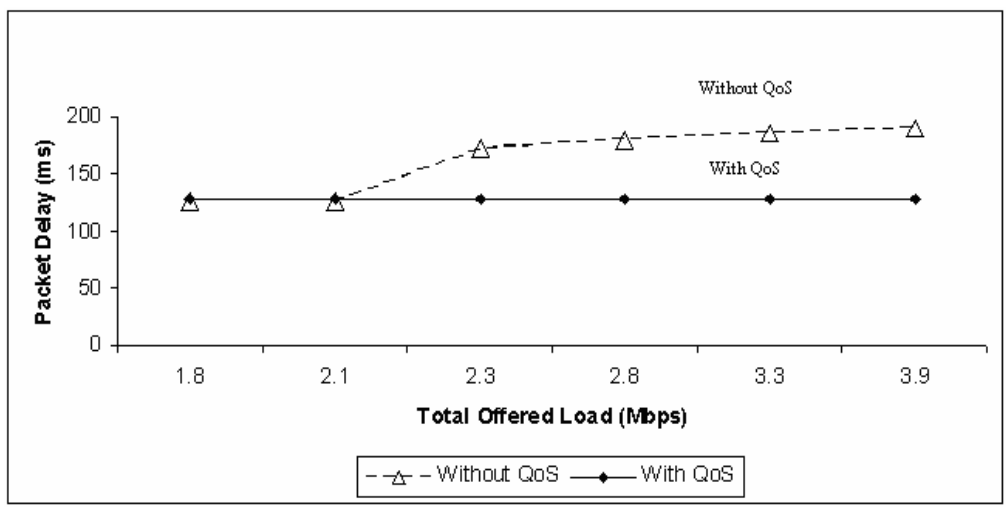

Figure 4.5(a): Packet delay vs. total offered load for AF traffic

Figure 4.5(a) shows that, as the total offered load at the bottleneck link increases, the packet delay at AF queue also increases, when QoS is not enabled in the bottleneck link. On the other hand, enabling QoS at the bottleneck link, the packet delay at AF queue remains same, though the total offered load at the bottleneck link increases beyond the available (constrained) bandwidth at the bottleneck link.

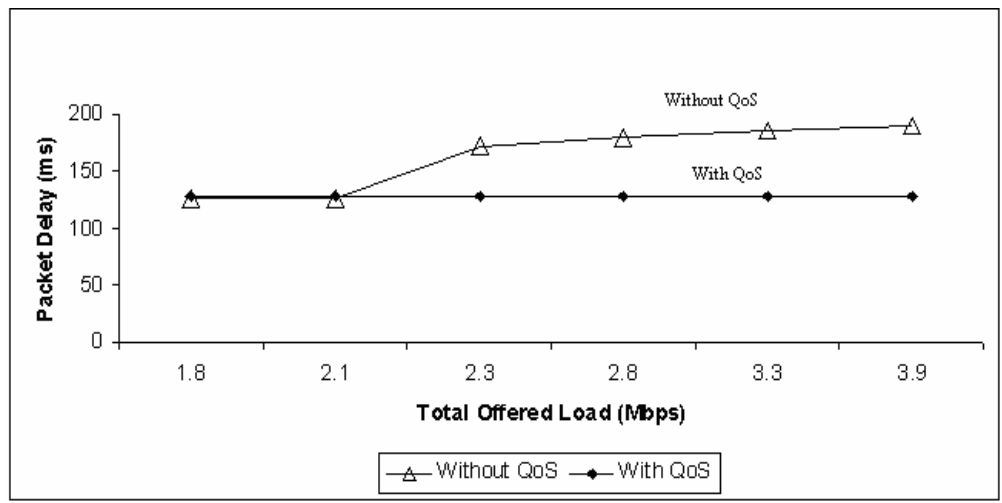

Figure 4.5(b): Packet delay vs. total offered load for EF traffic

Figure 4.5(b) shows that, as the total offered load at the bottleneck link increases, the packet delay at EF queue also increases, when QoS is not enabled in the bottleneck link. On the other hand, enabling QoS at the bottleneck link, the packet delay at EF queue remains same, though the total offered load at the bottleneck link increases beyond the available (constrained) bandwidth at the bottleneck link.

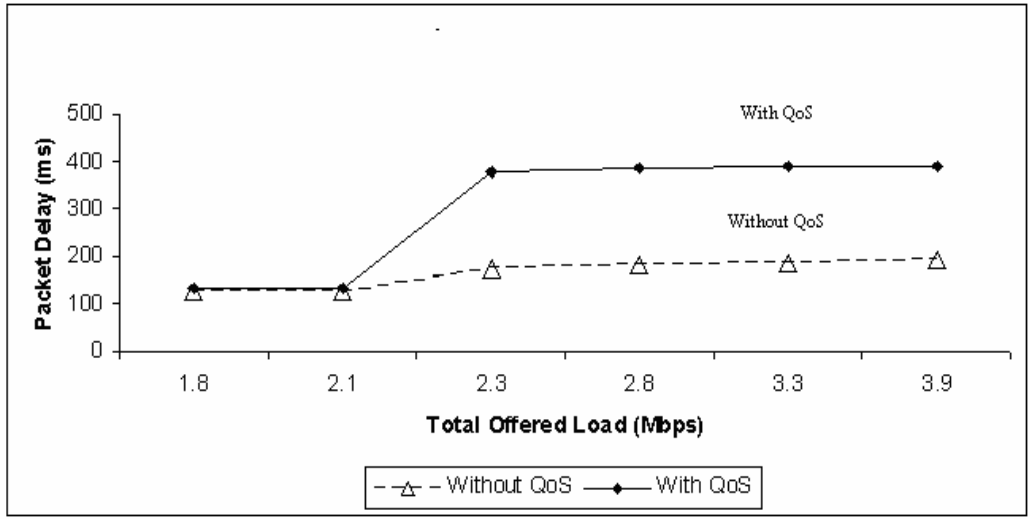

Figure 4.5(c): Packet delay vs. total offered load for BE traffic

Figure 4.5(c) shows that, as the total offered load at the bottleneck link increases, the packet delay at BE queue also increases, when QoS is not enabled in the bottleneck link. But enabling QoS at the bottleneck link, the packet delay at BE queue increases more rapidly when the total offered load are increasing beyond the available(constrained) bandwidth at the bottleneck link.

It is seen from Figures 4.5 (a, b and c) that the WRR scheduler ensures QoS to limit packet delay for each of the service until the total offered load is up to the bottleneck limit of 2.1 Mbps. Thereafter at higher total offered load the AF and EF packets continue to get guarantee of low packet delay whereas BE traffic gets affected. When total offered load exceeds the bottleneck bandwidth the priorities for AF and EF classes assure higher QoS for the classes repetitively where as the BE traffic class has higher packet delay.

### 4.2.3 Extension of the proposed model to multiple network domains

Considering multiuse of the resources, one can assume several sessions using the limited resources of the network. The approach considered here assumes two or more independent domains accessing the network with aggregated higher bandwidth and more resources. Such an approach also leads to a scaling up path to the activity. The mentioned simulated network topology has been enhanced for multiple domains of source traffic sharing common bottleneck link bandwidth [Figure 4.6].

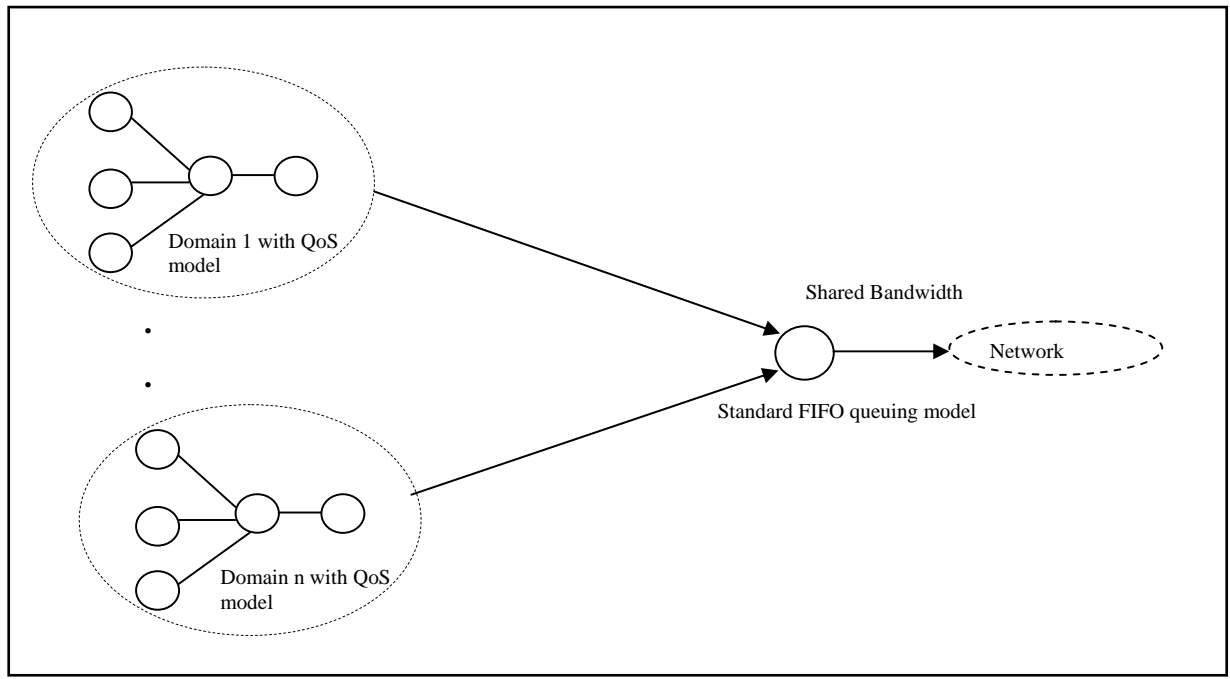

Figure 4.6: Simulated network topology for multiple domains

The proposed QoS model has been implemented in each domain. Standard FIFO scheduling has been implemented to serve multiple domains with shared link. Packet loss and packet delay statistics are observed at the receiver side. The AF, EF and BE traffic considered here are aggregated traffic. The simulation results have given the same nature as with single domain. If load increases the packet loss and delay increase at BE queues assuring the quality of AF and EF traffic.

To look at the scalability aspect, the proposed QoS model has been simulated with 4 domains. The shared common bottleneck link is 8.4 Mbps (MAX). Each domain gets 2.1 Mbps bandwidth. The proposed QoS model has been implemented in each domain. Standard FIFO queuing model has been considered at the shared link.

The packet loss and packet delay statistics are taken at the receiver side. The offered load at the link are increased by increasing the BE traffic. The results are shown in the Figures 4.7 (a, b and c) and Figures 4.8 (a, b and c). The X axis represents the total offered load at the shared link. It includes the aggregated AF, EF and BE traffic. The Y axis represents the packet loss [Figures 4.7 (a, b and c)] or packet delay [Figures 4.7 (a, b and c)] of AF, EF and BE traffic.

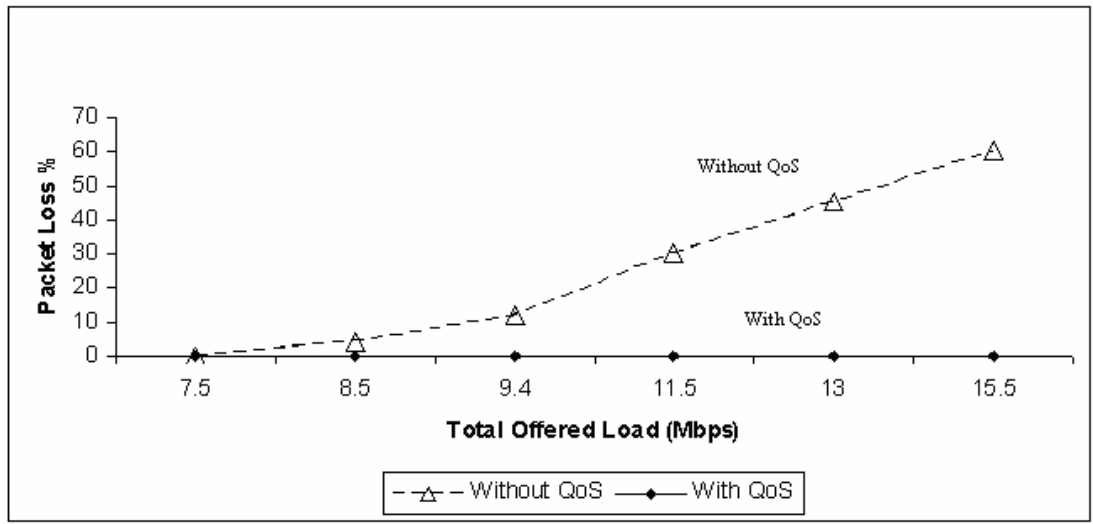

Figure 4.7(a): Packet vs. total offered load loss for AF traffic

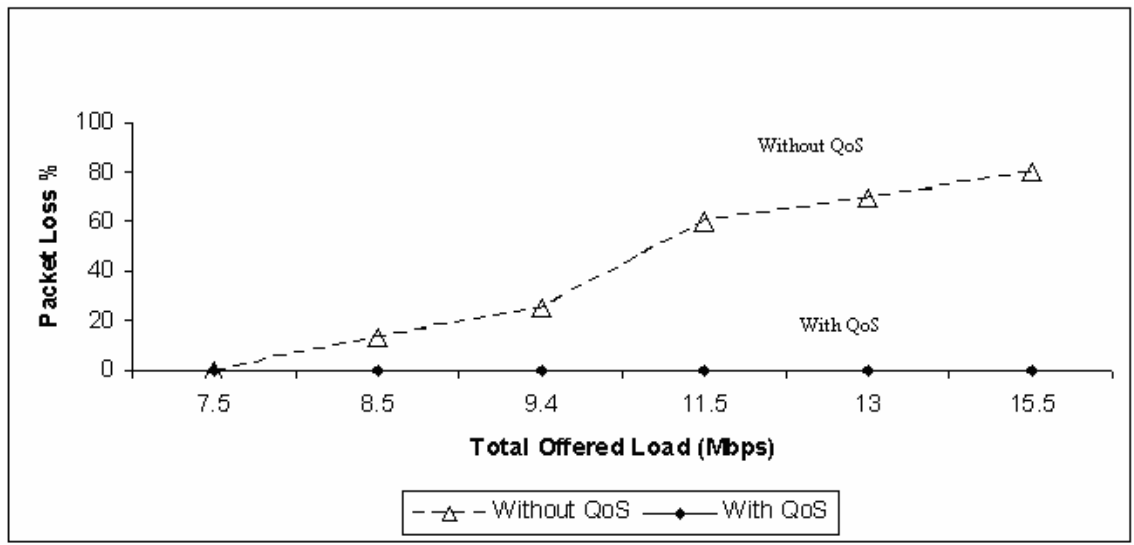

Figure 4.7(b): Packet loss vs. total offered load for EF traffic

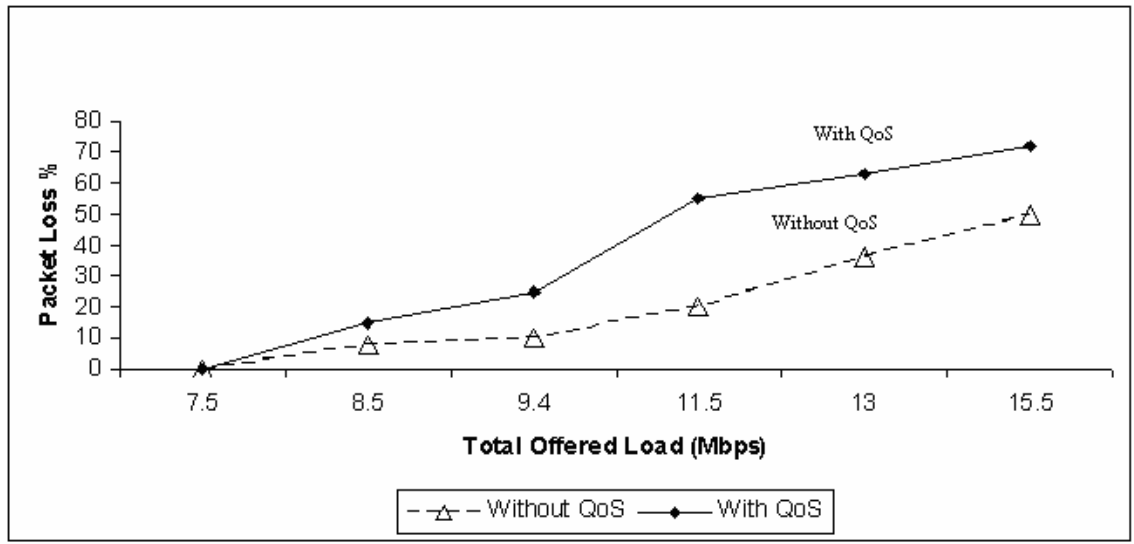

Figure 4.7(c): Packet loss vs. total offered load for BE traffic

Figures 4.7 (a, b and c) show that while implementing QoS scheduler, the AF and EF traffic packet loss remains at 0% though the total load is increasing. According to the scheduling algorithm the AF and EF traffic are getting priority over BE traffic. So, the packet loss for AF and EF traffic is 0% whereas packet loss at BE traffic increases with increase in load.

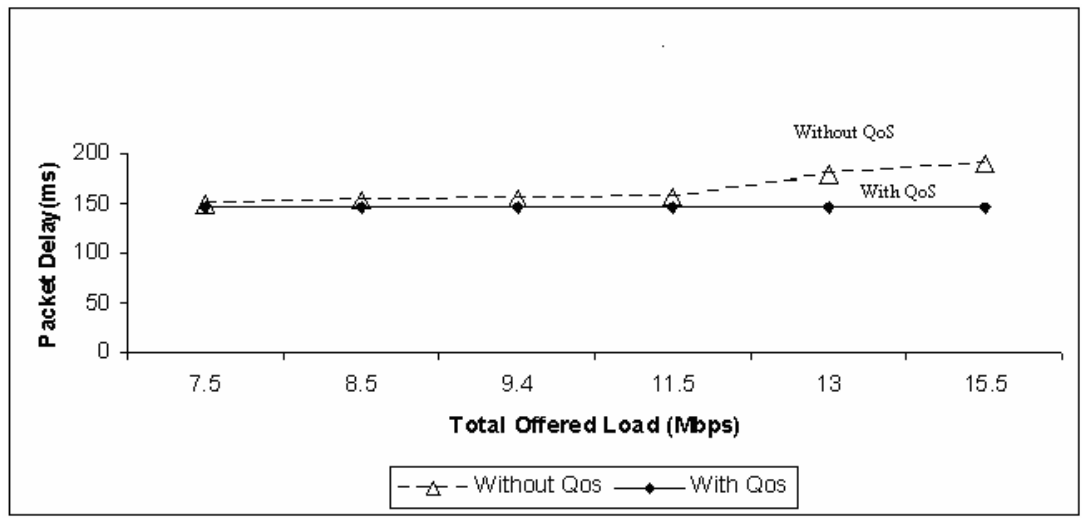

Figure 4.8(a): Packet delay vs. total offered load for AF traffic

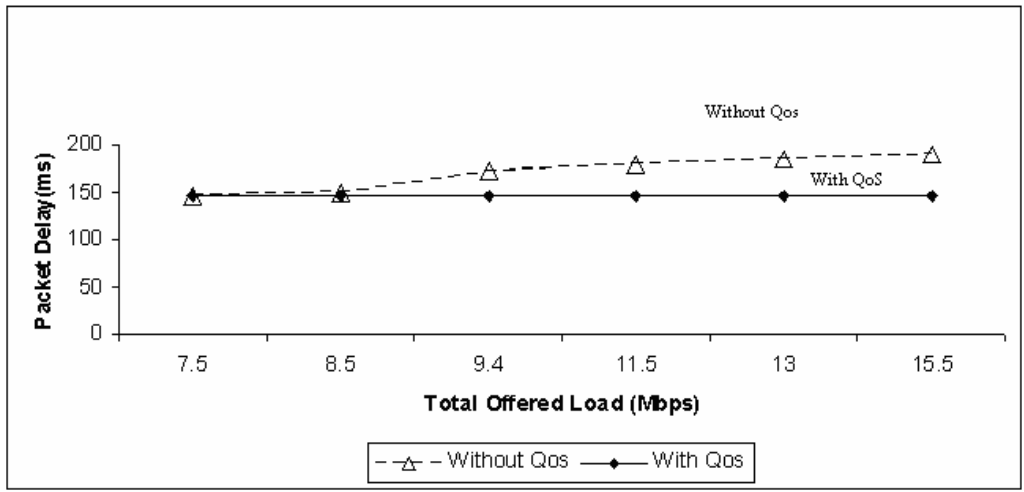

Figure 4.8(b): Packet delay vs. total offered load for EF traffic

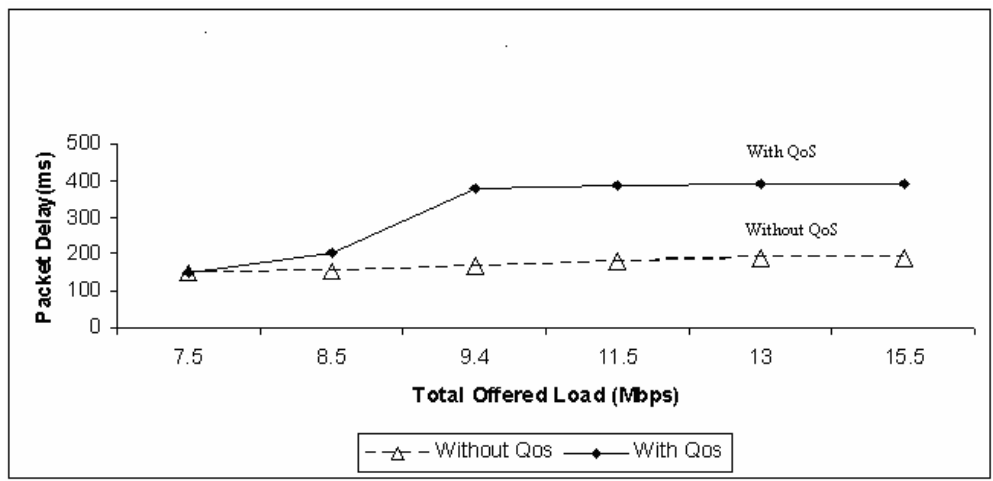

Figure 4.8(c): Packet delay vs. total offered load for BE traffic

Figures 4.8 (a, b and c) show that while implementing QoS scheduler, the AF and EF traffic packet delay remains at certain value though the total load is increasing. According to the scheduling algorithm the AF and EF traffic are getting priority over BE traffic. So, the packet delay for AF and EF traffic remains same whereas packet delay at BE traffic increases with increase in load.

It is seen from the above results that the individual WRR scheduler at each domain provides adequate QoS to limit packet loss and delay for each of the service until the total offered load is up to the bottleneck limit of 8.4 Mbps for the aggregates of all domains. Thereafter at higher total offered load the AF and EF packets continue to get guarantee of low packet loss and delay whereas BE traffic gets affected. When total offered load of the aggregate exceeds the bottleneck bandwidth the priorities for AF and EF classes assure higher QoS for the classes repetitively where as the BE traffic class has higher packet loss and delay.

The proposed QoS model is for providing quality assurance for real-time interactive traffic such as voice and video. It is not dealing with QoS for BE service. To provide fairness to BE traffic some congestion control mechanism may be implemented to BE queue which is not in the scope of this thesis work. From the simulation result it is clear that with this QoS scheduling model the packet loss and packet delay for real-time voice and video traffic are remaining at minimum value though the total offered load of the link is increasing beyond the available bottleneck bandwidth.

## 4.3 Real-time VBR service rate tuning parameter for sensitivity analysis of AF traffic

As described before real-time VBR (VBR-rt, generally video packets) traffic is considered for AF queue. Such traffic may exhibit packet size and data-rate variability and peak data rate may occur at any time. So, sudden packet drop may occur at AF queue even in presence of QoS.

Here a brief discussion on sensitivity analysis of AF traffic and some simulated results are presented. As mentioned earlier in section 4.2, the assigned WRR queue weight or service rate for different classes is as following:

Let, $qw_{AF}$ = assigned service rate for AF traffic.

$P_{AF}$ = EF traffic priority = 2.

$idr_{AF}$ = maximum input data rate for AF traffic.

$N_{AF}$ = number of sessions for AF traffic.

And now, let

$K = qw_{AF}$ / Computed AF queue service rate weightage.   (7)

K is multiplying factor which may be considered to tune the service rate of AF queue. If the assigned service rate weightage for AF traffic is less than computed AF queue service rate weightage then K is less than 1 else K is greater than 1. One can easily infer that if K <1, the assigned service rate will be less than the computed AF service rate and these may lead to more packet drops.

VBR-rt traffic has some peak for which high service rate for the AF queue will be useful to avoid packet drops. However since these peaks do not occur every time, it will be interesting to consider the influence of using lower than computed service rates for the AF queue.

K can also be seen as a service rate tuning parameter for the AF queue. It is useful in controlling the bandwidth assignment for the VBR-rt traffic. In practice, administrators and system implementers would like to economic by allotting lesser than computed service rate for this traffic as it may be more beneficial to the Best Effort traffic and other traffic. Therefore we analyze the sensitivity of the performance of the VBR-rt traffic between the end stations once the value of K, the tuning parameter is raised. Low K implies more bandwidth to the other traffic and high K reduces other traffic fairness. So K can also be considered in terms of a parameter which decides how much data rate can AF queue handle with no or negligible loss of packets.

The following graphs present the packet loss and delay for different value of K varying from 0.4 to 1.2. Figure-4.8 and Figure-4.9 show the % packet loss and packet delay respectively for VBR-rt traffic in the AF queue versus the total normalized offered load. The concept of total normalized load is as follows,
*Total normalized offered load = Offered traffic rate/Total link bandwidth.*
This is independent of the link bandwidth and can be used for any available access link bandwidth used in practice. In this case, 2.1 Mbps has been used as total link bandwidth that is available.

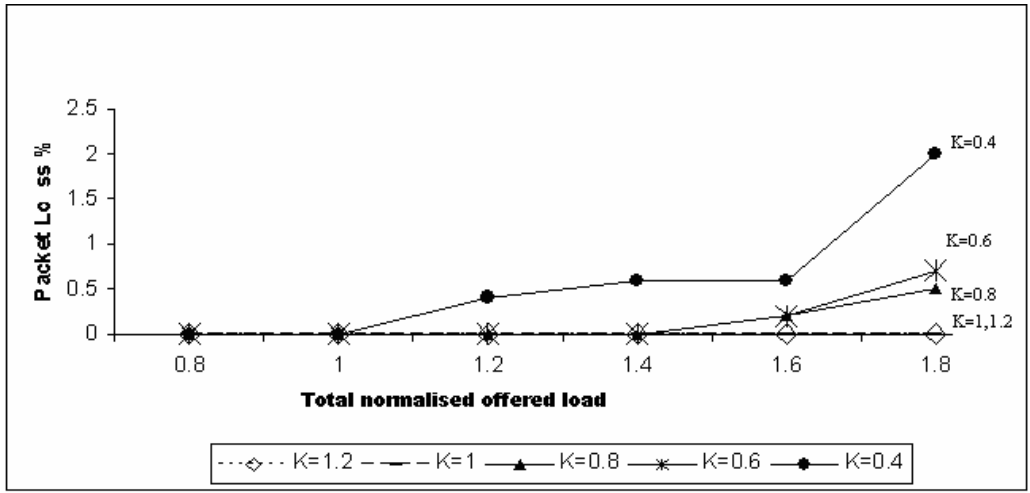

Figure 4.9: Packet loss vs. total normalized offered load for different service rate multiplying factor (K) for AF traffic

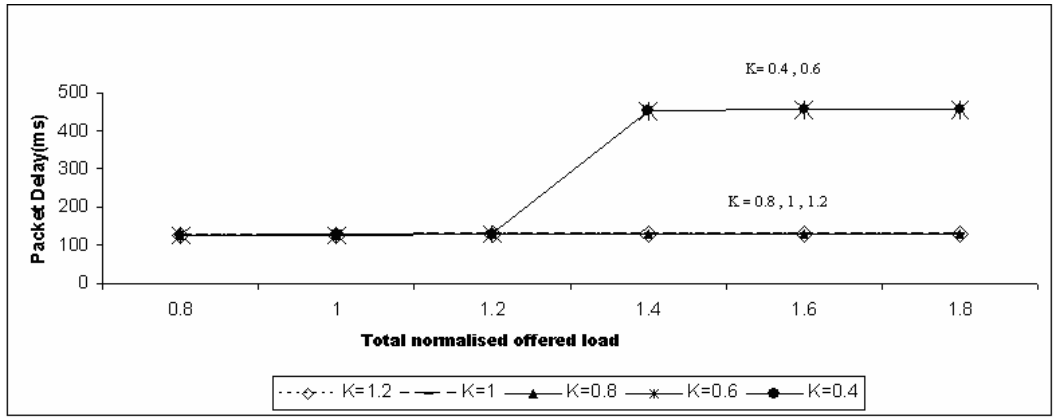

Figure 4.10: Packet delay vs. total normalized offered load for different service rate multiplying factor (K) for AF traffic

Figure 4.9 and Figure 4.10 show that for K = 0.4 the packet loss is 0% and delay $< =130$, for total offered load 2.1 Mbps. So when the total offered load is within available bandwidth to the link, the allocated service rate (queue weightage) for AF traffic can be brought down up to 2/5 times of the computed service rate, with same assured AF traffic quality of service. If the total offered load is increased beyond the available bandwidth, the allocated service rate of AF traffic can be brought down up to 4/5 times of the computed service rate.

This sensitivity analysis now helps to decide how much variation of allocated service rate can be made without compromising the quality of service to the real-time traffic. The advantage is that one can now take 2.5 times the no of VBR-rt sessions in the same computed AF service rate without reconfiguring the queues. Alternatively one can expect better performance to the other traffic even when such allocation is made.

Several experiments also have been conducted in the laboratory implementing proposed QoS model. The next chapter presents the experimental setup and the discussion on the results recorded from those experiments.

# Chapter 5

# Experimental Study and Analysis on a Test-bed

In chapter 4, a DiffServ based traffic engineering model for real-time interactive traffic in IP networks has been proposed and its performance analysis through the simulation results has been discussed. The proposed QoS model has been implemented in the experimental setup in the laboratory. Several experiments have been conducted and the results are recorded to validate the proposed QoS model. In this chapter, the experimental validation of proposed model on a setup made in a typical laboratory test-bed environment is presented. In simulation program the real-time and non real-time traffic were generated by NS2 simulator with variable bit rate and constant bit rate traffic sources. In experimental setup a video conferencing test tool application has been developed using Java packages to generate video and voice streams with different codec. The characteristics of different codec also have been studied to get the parameter, such as, input data rate and packet size of the traffic, pertaining to the traffic engineering model.

The experimental setup made in the laboratory is almost similar to a real-time network. First the actual network scenario will be discussed in the following section and then the laboratory setup and experimental results will be described.

## 5.1 Differentiated Service based IP network scenario

The networks are the collections of routers connected to each other. In differentiated service based IP networks routers used may be divided into two categories such as Core routers and Edge routers [Figure 5.1].

Core routers are routers which are only connected to other routers, not hosts. Core routers handle the aggregate flows. The important responsibilities of core routers are:

• Examining incoming packets for the code point marking done on the packet by the edge routers.

• Forwarding incoming packets into appropriate high bandwidth links according to their markings.

Edge routers or access routers are used between the hosts and the core network. The main features of edge routers are:

- Marking packets with a code point that reflects the desired level of service.
- Scheduling and controlling traffic to ensure QoS.
- Ensuring that user traffic adheres to its policy specifications, by shaping and policing traffic.

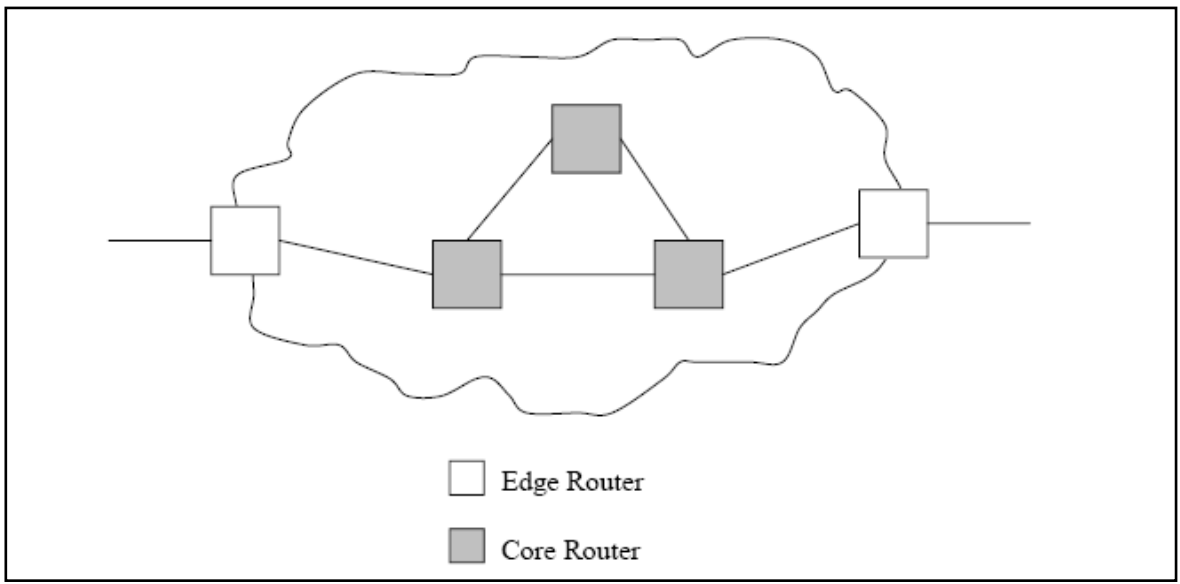

Figure 5.1: Network devices for DiffServ domain

The traffic sources and sinks are connected to edge routers. The routers are programmed to provide QoS to the traffic.

In the suggested QoS provisioning model described in this thesis, the QoS implementation has been divided in 3 steps according to differentiated service. These are,

1. Marking of different traffic,
2. Classifying of different traffic and
3. Scheduling of different traffic.

The basic formulation has already been described in the previous chapters. Marking is done at the ingress of the edge router of the network at source side. Classifying and Scheduling are done at the egress of the edge routers. Figure 5.2 is the block diagram of the actual network like experimental setup emulated in the laboratory. The source is located is Network 1 and the sink is in Network 2. The same QoS model has been replicated to the sink side also to provide bidirectional QoS enabled communication.

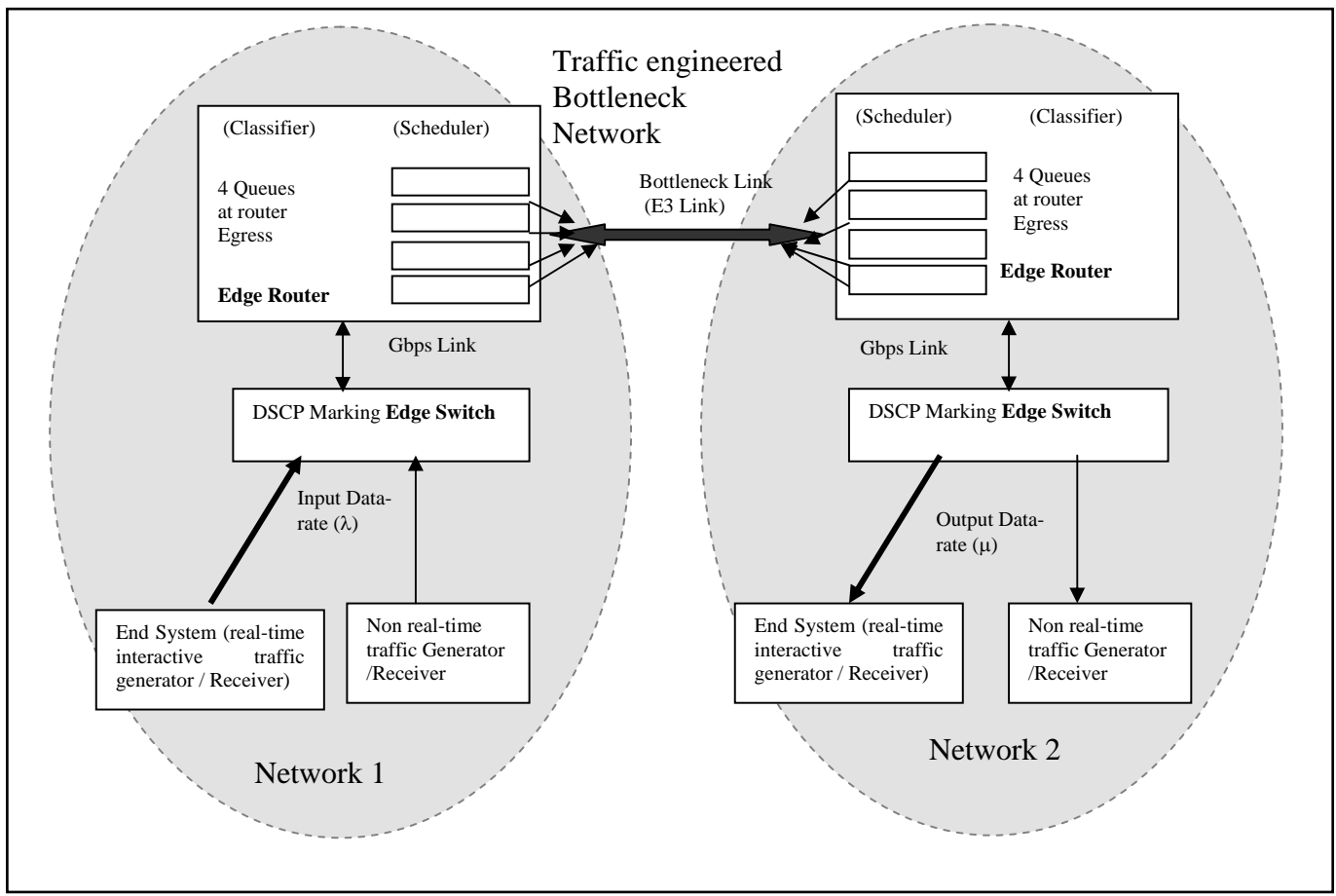

Figure 5.2: Block diagram of local test bed at laboratory

## 5.2 Experimental setup in the laboratory

To study the network performance with this QoS model a test setup has been emulated in the laboratory. The setup reflects a simple yet similar scenario in an actual network environment.

The layer three switches Nortel 3510 [27] is used as edge router which is connected to the different traffic generator sources and marks the packets of different real-time traffic with different DSCP value. Nortel 3510 switch provides enough bandwidth of 1 Gbps to the edge router.

The Nortel switch is connected to Juniper M7i [28] router, which is the edge routers and classifies and schedules the incoming marked traffic from the Nortel switch. Juniper M7i router in its basic configuration supports 4 logical queues with 100 ms delay buffer. The queues are for Assured Forwarding, Expedited Forwarding, Best Effort and Network Control. By default Best Effort queue (95%) and Network Control queue (5%) are defined.

A test bed has been setup in the laboratory with two networks and a bottleneck link as shown in Figure 5.3. The bandwidth of the E3 Link of Juniper M7i router can be sub rated between 358 Kbps and 34Mbps to create bottleneck link. It has been sub-rated to 2.1 Mbps to create a practical available condition of 2.048 Mbps or E1 link. The egress port E3 has been configured for QoS as described earlier. The same QoS model has been implemented to both the network to have both way communications with QoS.

A Java based videoconferencing application (Using Java media Framework) [29] at systems A and B has been designed for sending and receiving video and voice streams simultaneously. With this application different video and voice streams using different codecs can be generated. When voice and video packets come to the Nortel switch, they are marked with DSCP value by the Nortel switch.

Other traffic in the network is modeled through a cross traffic generator. The cross traffic has been generated through traffic generator (Distributed Internet Traffic Generator) [30] to choke the bottleneck link. There may be various types of traffic characteristics, such as constant packet size, variable packet size, constant bit rate, and variable bit rate.

The Nortel switch does not mark the cross traffic and leaves it at its default DSCP value of 0 (Best Effort). This adequately represents a general usage scenario. The volume of traffic generated can be controlled from the program.

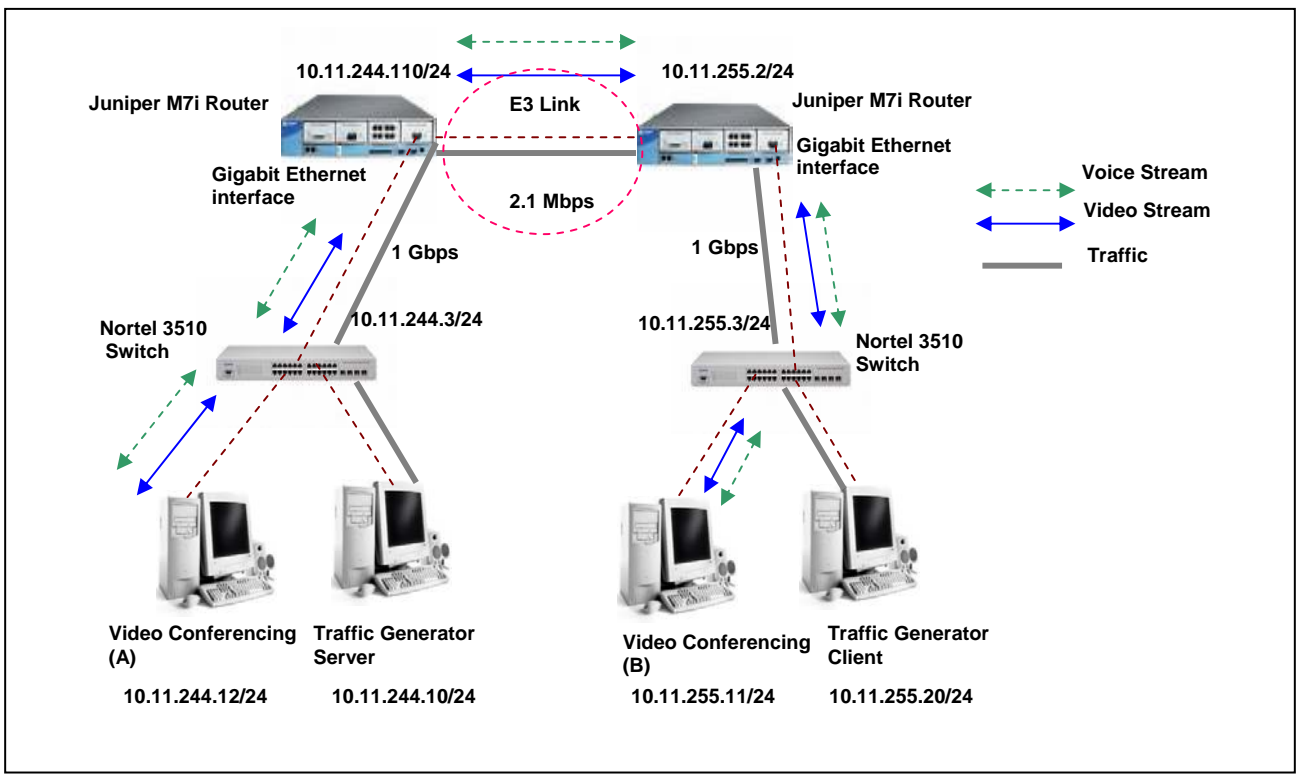

Figure 5.3: Experimental setup in the laboratory

When the video, voice and the traffic generated from traffic generator come to the Juniper M7i router it is distributed among 3 queues, assured forwarding queue for video, expedited forwarding queue for voice and best effort queue for the traffic generated from traffic generator.

A network analyzer Wire-shark [31] has been used to measure the packet loss and delay by audio and video traffic at the end system (application level). This way the characteristics of the traffic as seen in the end system is better measured and reported. The end-to-end QoS characteristics are reported henceforth.

### 5.2.1 The traffic engineering model supported in network test-bed

Juniper routers implement QoS model provisioning mechanisms in a specific manner. These are typically an implementation dependent factor. But this influences the overall implementation of QoS and end-to-end performance. The provisioning of bandwidth and buffer vis-à-vis traffic types is described from the juniper documentation [33]. These are:
- 4 logical queues in a typical basic configuration.
- First in First out (FIFO), Priority Scheduling (PS), Round Robin (RR), Weighted Round Robin (WRR) schedulers are available to serve the queues.
- Total delay buffer available at Juniper router is 100 ms.
- Link rate at E3 link can be sub-rated from min 358 Kbps to 34 Mbps. 2.1 Mbps has been taken for the experiment, as it is the available bandwidth of the most real situation. This is based on a leaky bucket traffic shaping model to limit the outgoing bandwidth.

Nortel 3510 switch is layer 3 switches with 8 logical queues and some QoS configuration features. The switch can be configured for marking, shaping and scheduling traffic [27]. For the experiments the following configurations are made in the Nortel 3510 switch,
- Port based packet marking is enabled at the switch. It marks the video and voice traffic as AF (12) and EF (46) DSCP value respectively. For the other traffic, it leaves their default DSCP value as 0.
- The link between Nortel switch to Juniper router is 1 Gbps.

The video conferencing equipments like JMF based video conferencing application [30], Tandberg 6000MXP [32] and Robotic surveillance camera [34] are used as end system which generates the video and voice traffic with different codecs. Link bandwidth from end system to Nortel Switch is 1 Gbps. No Scheduler is enabled at the end system. All scheduling is carried out at edge router only.

### 5.2.2 Assumptions for measurement

Some assumptions have been made for the experiments done. These are:
- The input, output data rate, packet loss and delay are taken from the end system only – not at the router side. This gives a proper estimate of end-to-end QoS for the application.
- The packet loss measured includes the loss due to entire network – not only for a particular forwarding queues corresponding to a particular traffic.
- The packet delay measured includes the loss due to entire network – not only for a particular forwarding queues corresponding to a particular traffic.
- The delay depends on various system clocks present in the network. For the following experiment all the clocks are made synchronized manually as much as possible. Better measurement may be achieved by using a NTP server for synchronizing system clocks.

### 5.2.3 Real-time interactive transmission performance requirement

To study and test the network performance with the proposed QoS model, the results need to be compared with a standard network performance requirement. Here is a typical video conferencing requirement is presented in the Table 5.1[17].

Table 5.1 Video conferencing requirement

| Real time traffic | Packet Loss | Packet Delay | Jitter |
| --- | --- | --- | --- |
| Voice | Reasonably resilient to packet loss | <150 ms | <30ms |
| Video | Low packet loss is primary requirement | <150 ms | less concern |

The QoS provisioning model considered here is concentrating on improvement of Packet loss and delay performance of real-time interactive traffic, controlling the QoS parameter at network side. Jitter or delay variation is mostly related to application side as it relates to setting up parameters in the end system application.

## 5.2.4 Traffic source characterization for different traffic sources

The real-time traffic can be classified between Constant Bit Rate real-time (CBR-rt) traffic and Variable Bit Rate real-time (VBR-rt) traffic. The example of CBR-rt traffic is audio traffic such as MPEG audio, ulaw, GSM codec. The example of VBR-rt traffic is video traffic such as H263, JPEG-RTP codec. On the basis of experiment carried out in the laboratory some characteristics of codec are presented here.

A video conferencing application has been developed using JMF for generating real-time voice and video traffic. The voice and video traffic are consists of the codec mentioned above. The sender application takes the live picture and audio and sends to the receiver application through the QoS enabled network device. The data rate and packet size of the generated traffic are measured by wire shark network analyzer at the sender side where the real-time traffic are generated by JMF application. The measured parameters are taken to compute the QoS scheduling parameter.

The characterization of video and voice traffic generated by JMF application is described in the following subsections.

**Characteristics (experimental observation) of video codecs**

a) Max Packet size observed = $pktsz_{AF\,(max)}$.

   Ex: $pktsz_{AF\,(max)}$ observed for H263 = 1038 Bytes.

   $pktsz_{AF\,(max)}$ observed for jpeg-rtp = 1022 Bytes.

   Figure 5.4 And Figure 5.5 show packet size distributions for H263 and JPEG-RTP codec respectively.

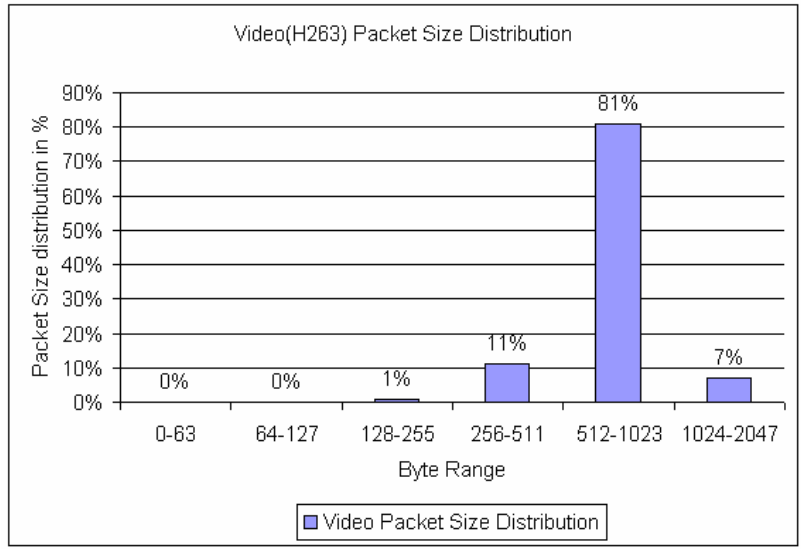

Figure 5.4: Packet size distribution of H263 codec with RTP (using Wire-shark)

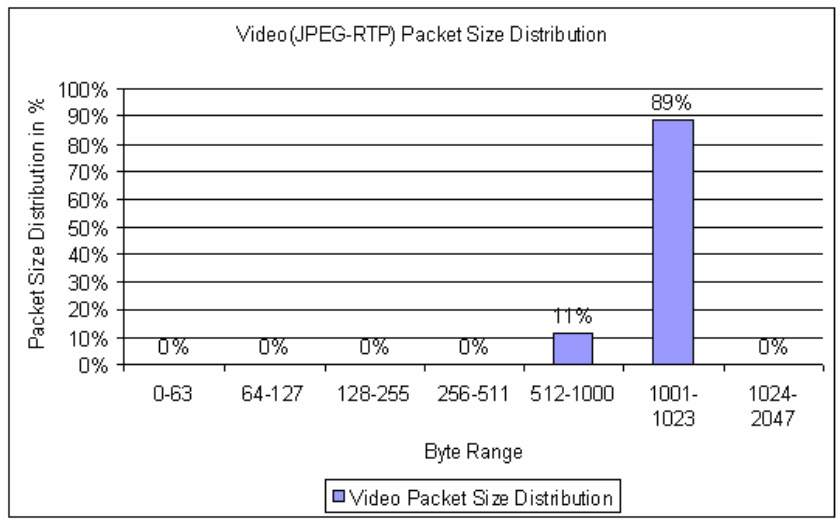

Figure 5.5: Packet size distribution of JPEG-RTP codec with RTP (using Wire-shark)

b) Average input data rate = $\lambda_{A.F.}$

   Ex: For H263 $\lambda_{AF}$ observed is 328 Kbps.

   For JPEG-RTP $\lambda_{AF}$ observed is 1.05 Mbps.

c) Peak input data rate = $idr_{AF.}$

   Ex: For H263 $idr_{AF}$ observed is 840 Kbps.

   For JPEG-RTP $idr_{AF}$ observed is 1.4 Mbps.

Figure 5.6 and Figure 5.7 show average and peak data rate for H263 and JPEG-RTP codec respectively.

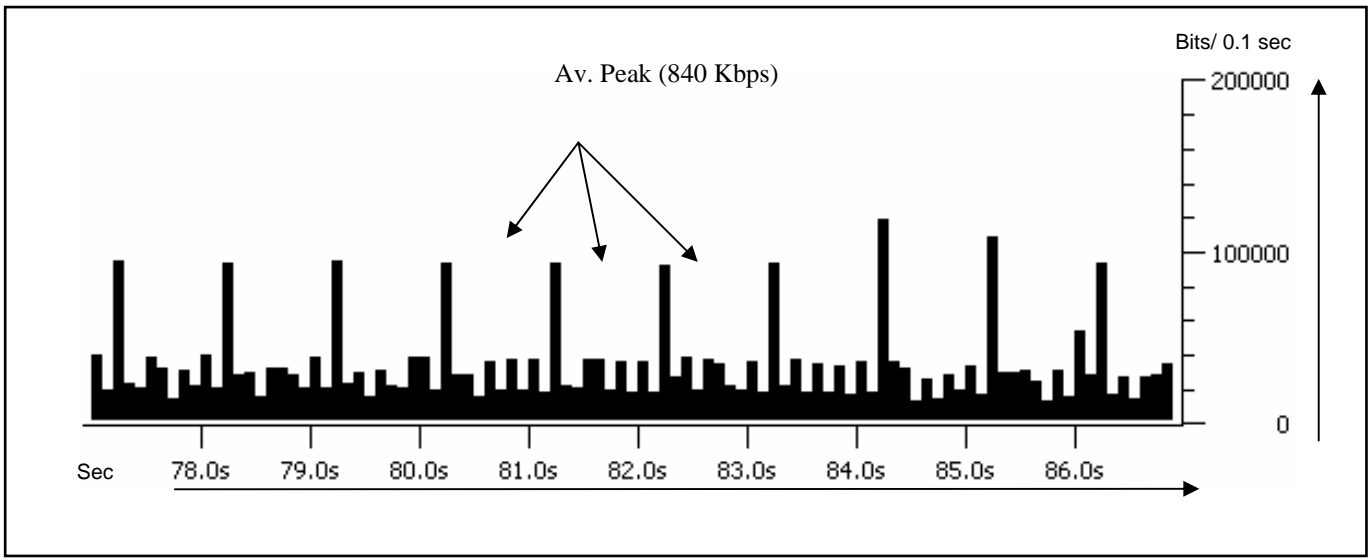

Figure 5.6: Data rate of H263 codec (using Wire-shark)

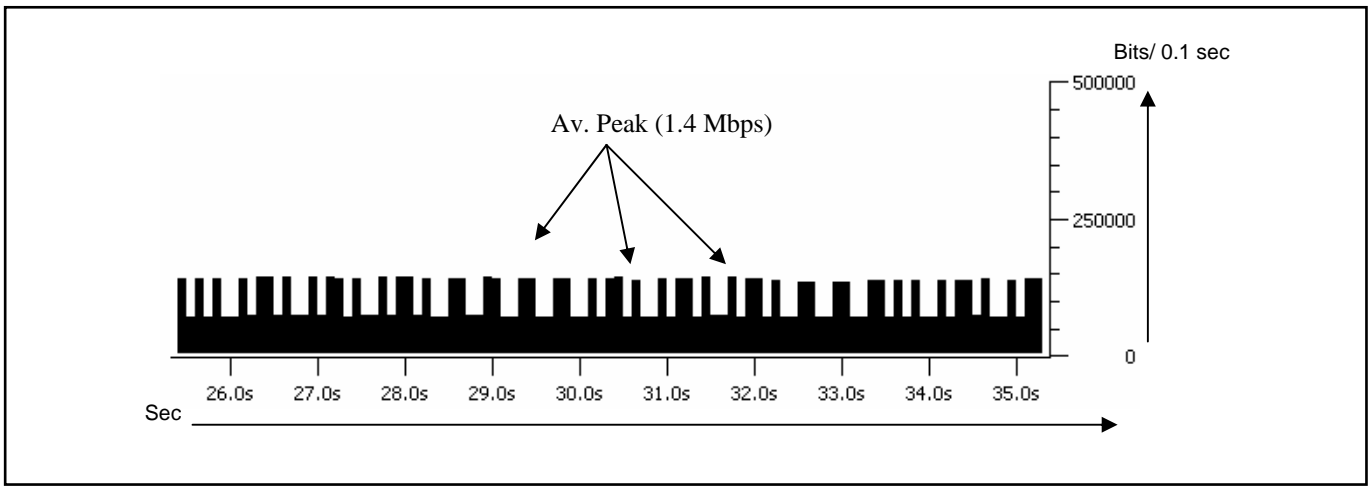

Figure 5.7: Data rate of JPEG-RTP codec (using Wire-shark)

**Characteristics (experimental observation) of voice codecs**

a) Average packet size = $pktsz_{EF}$.

  Ex: $pktsz_{EF}$ observed for MPEG audio = 1402 Bytes.

   $pktsz_{EF}$ observed for G723 = 102 Bytes.

   $pktsz_{EF}$ observed for GSM = 153 Bytes.

$pktsz_{EF}$ observed for μlaw= 534 Bytes.

$pktsz_{EF}$ observed for DVI-RTP= 298 Bytes.

Figure 5.8 shows average packet size distribution for different voice codecs.

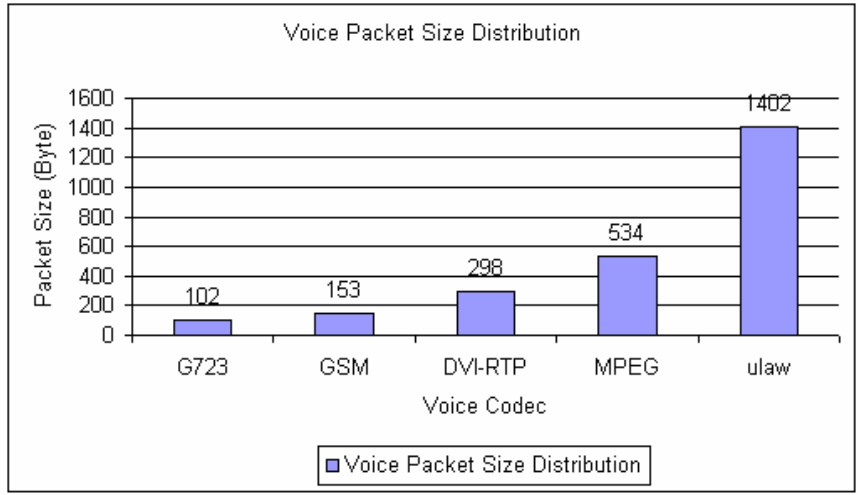

Figure 5.8: Packet size distributions for different voice codecs (using Wire-shark)

b) Average data rate = $\lambda_{EF}$ = $idr_{EF}$.

Ex: $idr_{EF}$ observed for MPEG audio = 64 Kbps.

$idr_{EF}$ observed for G723 = 14 Kbps.

$idr_{EF}$ observed for GSM = 20 Kbps.

$idr_{EF}$ observed for μlaw = 71 Kbps.

$idr_{EF}$ observed for DVI-RTP = 40 Kbps.

Figure 5.9 shows average data rate distributions for different voice codecs.

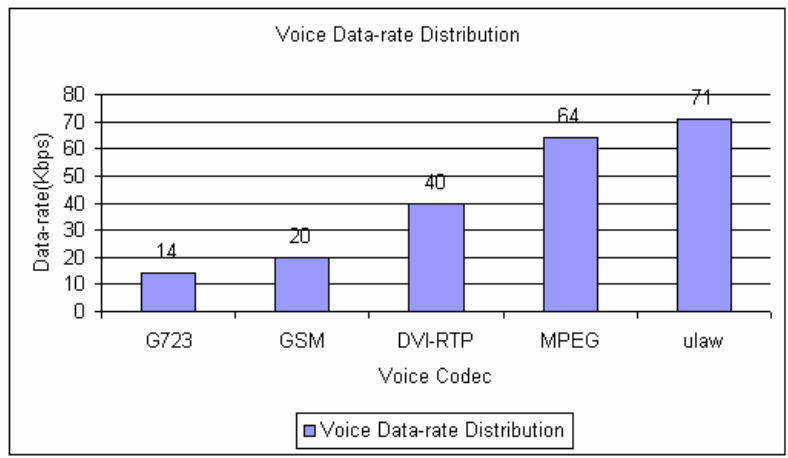

Figure 5.9: Data rate distribution for different voice codecs (using Wire-shark)

## 5.3 Recomputed QoS provisioning scheduling parameter in laboratory experiments

The proposed QoS scheduling algorithm presented in the previous chapter is dynamic in nature. The computed service rate is adaptive to the queue length of the traffic. It is often observed that several practical router implementations follow a scheduling mechanism which may be different from the one discussed earlier. Most of the implementations as in Cisco and Juniper routers have a fixed allocation of service rate weightage and policies. In such situations the computed service rate or queue weightage is not adaptive with change of the queue length rather it is the function of input data rates of different classes of traffic. Assigned WRR queue weight for different classes is as follows:

Let, $qw_{AF}$ = queue weightage or service rate for AF traffic.

$qw_{EF}$ = queue weightage or service rate for EF traffic.

$qw_{BE}$ = queue weightage or service rate for BE traffic.

$P_{AF}$ = EF traffic priority = 2.

$P_{EF}$ = EF traffic priority = 3.

$P_{BE}$ = EF traffic priority = 1.

$idr_{AF}$ = maximum input data rate for AF traffic.

$idr_{EF}$ = average input data rate for EF traffic.

$idr_{BE}$ = average input data rate for BE traffic.

$N_{AF}$ = number of sessions for AF traffic.

$N_{EF}$ = number of sessions for AF traffic.

TB = total available bandwidth at bottleneck link.

Now the computed service rate is as follows.

$qw_{AF} = [(idr_{AF} / TB)*100] * N_{AF} * P_{AF}.$      (8)

$qw_{EF} = [(idr_{EF} / TB)*100] * N_{EF} * P_{EF}.$      (9)

$qw_{BE} = [(idr_{BE} / TB)*100] * P_{BE}.$      (10)

## 5.4 Experimental results and analysis

Several sets of experiments have been conducted to observe the effect of proposed QoS model on the real-time traffic, increasing traffic load at bottleneck link bandwidth by increasing non real-time traffic rate at BE queues, creating the following scenarios,
(a) With implementing the proposed QoS model and
(b) Without implementing the proposed QoS.

The real-time traffic has been generated with a Java application test tool developed to generate video and voice packets simultaneously. Different experiments also have been arranged with different types of real-time sources such as Tandberg video conferencing equipment and a robotic surveillance camera. The results are presented in the following subsections.

## 5.4.1 Performance of a network with QoS traffic engineering applied for Java based application

A network end to end QoS performance testing tool has been developed by JMF application to generate video and voice packets with different codes. It supports H263 and JPEG-RTP for video packets and MPEG-I/II, GSM, DVI-RTP, G723 and μlaw for voice packets. The following figures show how the packet loss and packet delay of voice (EF) and video (AF) traffic at receiver side are affected in the presence or absence of proposed QoS model at the router with increasing traffic load at the link. Following results are of video H263 and voice MPEG-I/II. Single video and voice session are considered for the experiment. The total offered load at the bottleneck link is increased by increasing the non real-time BE traffic generated by DITG traffic generator. The BE traffic is assumed to be Poisson distributed.

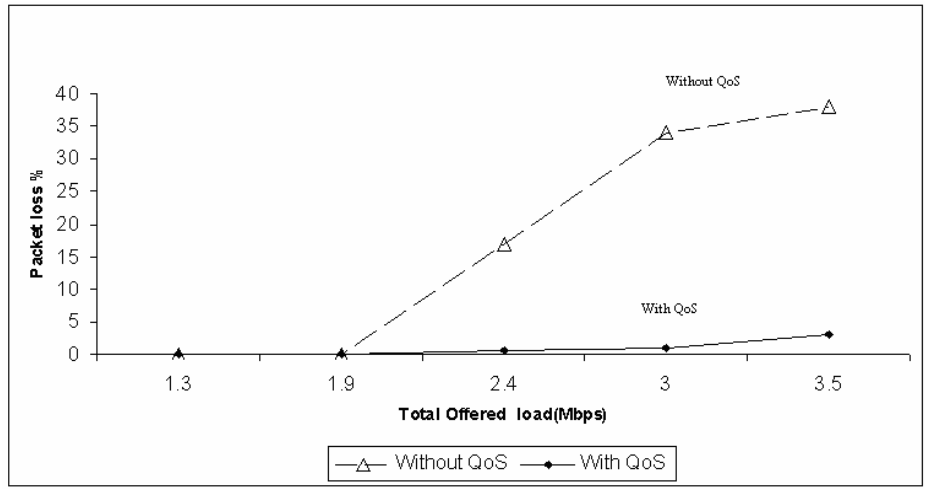

Figure 5.10(a): Packet loss vs. total offered load for AF traffic with bottleneck link bandwidth 2.1 Mbps and JMF as a real-time traffic source

Figure 5.10(a) shows that, as the total offered load at the bottleneck link increases, the packet loss at AF queue also increases rapidly, when QoS is not enabled in the bottleneck link. On the other hand, enabling QoS at the bottleneck link, the packet loss at AF queue remains at 0% while the total offered load is less than the available bandwidth at the bottleneck link (2.1 Mbps). The packet loss at AF queue increase very slowly while the total offered load at the bottleneck link increases beyond the available (constrained) bandwidth at the bottleneck link.

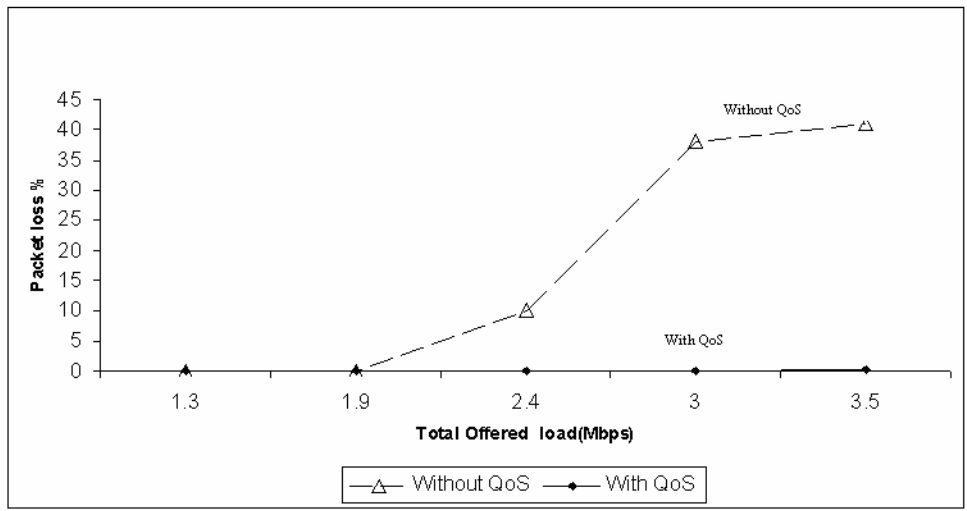

Figure 5.10(b): Packet loss vs. total offered load for EF traffic with bottleneck link bandwidth 2.1 Mbps and JMF as a real-time traffic source

Figure 5.10(b) shows that, as the total offered load at the bottleneck link increases, the packet loss at EF queue also increases rapidly, when QoS is not enabled in the bottleneck link. On the other hand, enabling QoS at the bottleneck link, the packet loss at EF queue remains at 0%, though the total offered load at the bottleneck link increases beyond the available (constrained) bandwidth at the bottleneck link.

Implementing the suggested QoS scheduler, the packet loss at AF and EF queues remains almost at 0% though the total offered load at bottleneck link is increasing. According to the proposed scheduling algorithm the AF and EF traffic are getting priority over BE traffic. So, the packet loss for AF and EF traffic is negligible whereas packet loss at BE traffic will increase with increase in offered load. Figure 5.10(a) and Figure 5.10(b) present only the real-time traffic.

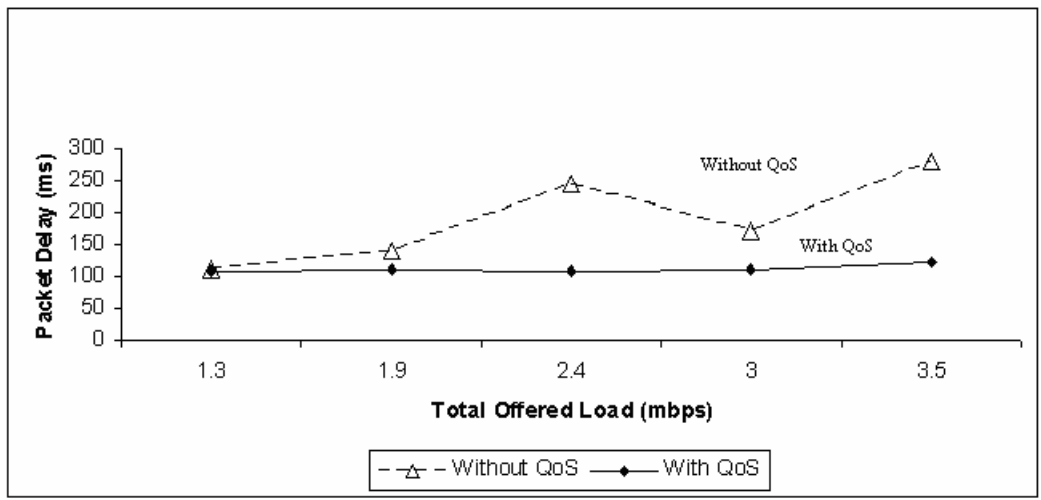

Figure 5.11(a): Packet delay vs. total offered load for AF traffic with bottleneck link bandwidth 2.1 Mbps and JMF as a real-time traffic source

Figure 5.11(a) shows that, as the total offered load at the bottleneck link increases, the packet delay at AF queue also increases, when QoS is not enabled in the bottleneck link. On the other hand, enabling QoS at the bottleneck link, the packet delay at AF queue remains almost same (less than 130 ms), though the total offered load at the bottleneck link increases beyond the available (constrained) bandwidth at the bottleneck link.

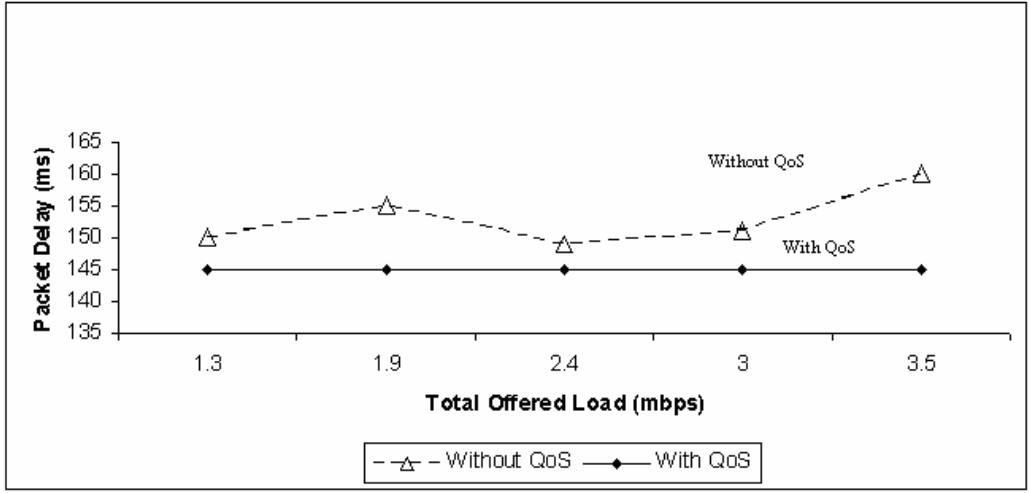

Figure 5.11(b): Packet delay vs. total offered load for EF traffic with bottleneck link bandwidth 2.1 Mbps and JMF as a real-time traffic source

Figure 5.11(b) shows that, as the total offered load at the bottleneck link increases, the packet delay at EF queue also increases, when QoS is not enabled in the bottleneck link. On the other hand, enabling QoS at the bottleneck link, the packet delay at EF queue remains almost same (less than 150), though the total offered load at the bottleneck link increases beyond the available (constrained) bandwidth at the bottleneck link.

Implementing proposed QoS scheduler, the packet delay at AF and EF queue remains at certain value though the total offered load to the bottleneck link is increasing. According to the scheduling algorithm the AF and EF traffic are getting priority over BE traffic. So, the packet delay for AF and EF traffic remains same whereas packet delay at BE traffic will increase with increase in offered load. Figure 5.11(a) and Figure 5.11(b) present only the real-time traffic.

It is seen from the above results that the proposed model provides QoS to limit packet loss and delay for each of the service as generated from the JMF test tool for real-time video and audio until the total offered load is up to the bottleneck limit of 2.1 Mbps. Thereafter at higher total offered load the AF and EF packets continue to get guarantee of low packet loss and delay whereas BE traffic gets affected. When total offered load exceeds the bottleneck bandwidth the priorities for AF and EF classes assure higher QoS for the classes repetitively where as the BE traffic class has higher packet loss and delay.

## 5.4.2 Performance of network with QoS traffic engineering for videoconferencing equipment

A set of trials have been conducted with Tandberg videoconferencing equipment. It generates video streams using codecs H261, H264 and voice codes of G723, MPEG codec. The following results are of video H261 and voice G723. These sets of experiments also reflect that implementing the proposed QoS provisioning mechanism improves the performance of real-time traffic. Single video and voice session are considered for the experiment. The total offered load at the bottleneck link is increased by increasing the non real-time BE traffic generated by DITG traffic generator. The BE traffic is assumed to be Poisson distributed.

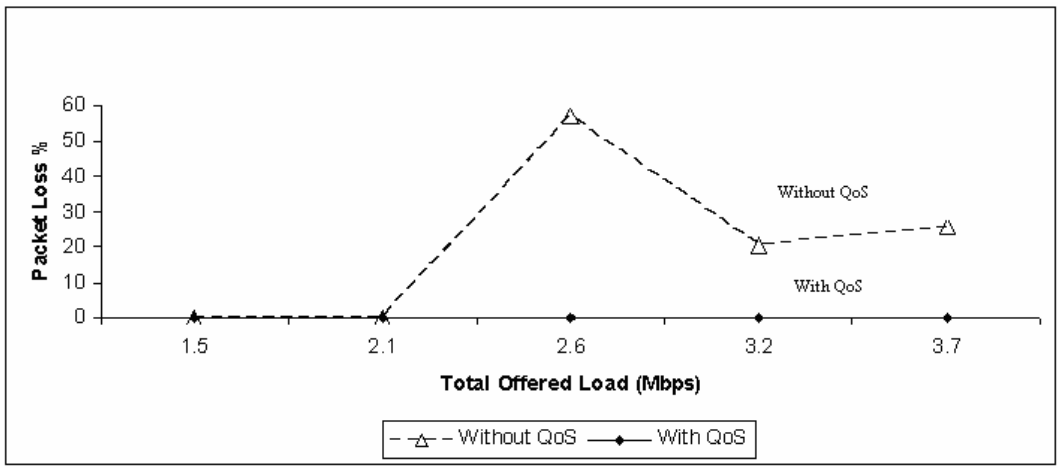

Figure 5.12(a): Packet loss vs. total offered load for AF traffic with bottleneck link bandwidth 2.1 Mbps and Tandberg system as a real-time traffic source

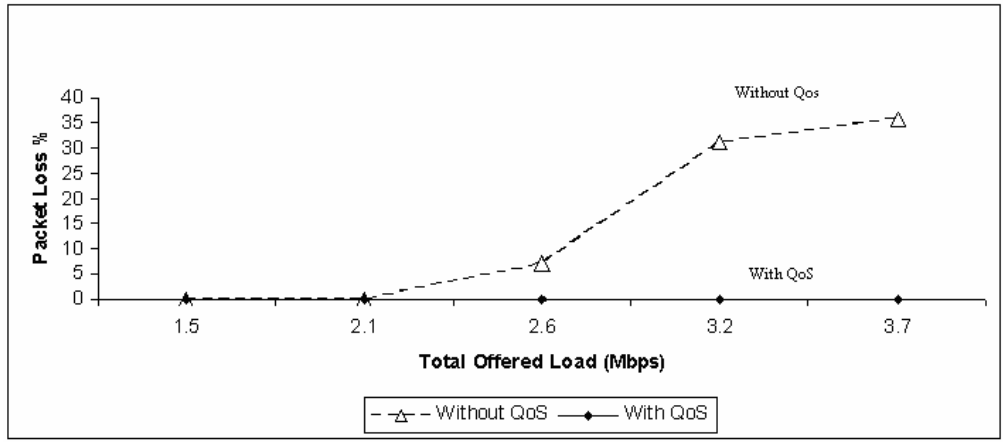

Figure 5.12 (b): Packet loss vs. total offered load for EF traffic with bottleneck link bandwidth 2.1 Mbps and Tandberg system as a real-time traffic source

Figure 5.12(a) and Figure 5.12(b) show that, as the total offered load at the bottleneck link increases, the packet loss at AF and EF queue also increases, when QoS is not enabled in the bottleneck link. On the other hand, enabling QoS at the bottleneck link, the packet loss at AF and EF queue remains at 0%, though the total offered load at the bottleneck link increases beyond the available (constrained) bandwidth at the bottleneck link. According to the proposed scheduling algorithm the AF and EF traffic are getting priority over BE traffic. So, the packet loss for AF and EF traffic is negligible whereas packet loss at BE traffic will increase with increase in offered load. Figure 5.12(a) and Figure 5.12(b) presents only the real-time traffic.

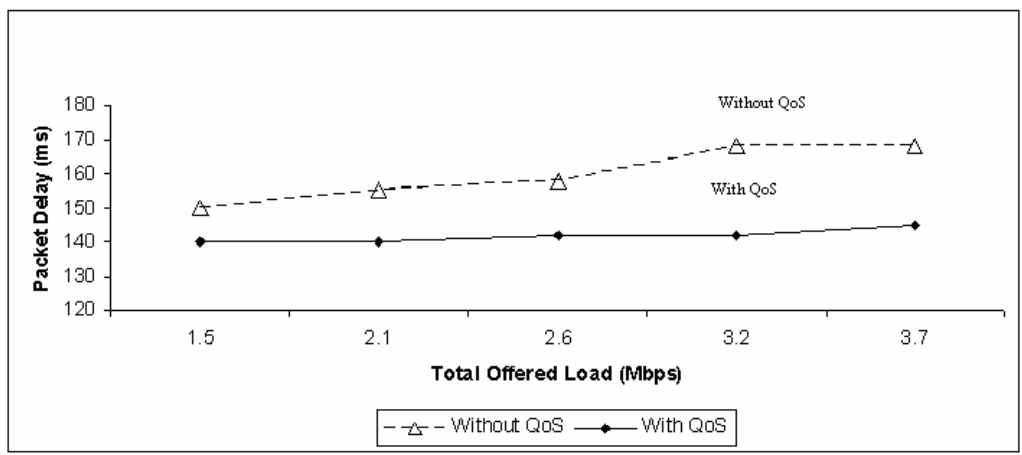

Figure 5.13(a): Packet delay vs. total offered load for AF traffic with bottleneck link bandwidth 2.1 Mbps and Tandberg system as a real-time traffic source

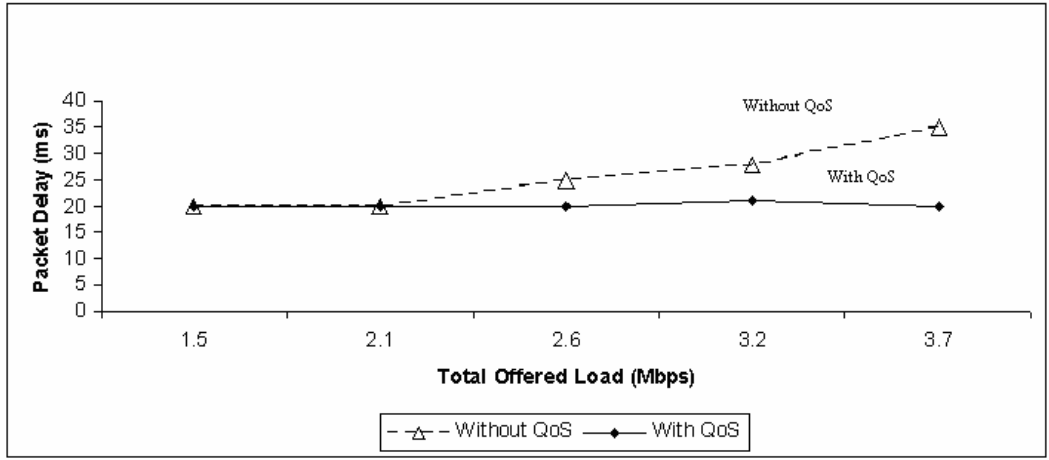

Figure 5.13(b): Packet delay vs. total offered load for EF traffic with bottleneck link bandwidth 2.1 Mbps and Tandberg system as a real-time traffic source

Figure 5.13(a) and Figure 5.13(b) show that, as the total offered load at the bottleneck link increases, the packet delay at AF and EF queues also increases, when QoS is not enabled in the bottleneck link. On the other hand, enabling QoS at the bottleneck link, the packet delay at AF and EF queue remains almost same, though the total offered load at the bottleneck link increases beyond the available (constrained) bandwidth at the bottleneck link. According to the proposed scheduling algorithm the AF and EF traffic are getting priority over BE traffic. So, the packet delay for AF and EF traffic remains almost same(less than 150 ms) whereas packet delays at BE traffic will increase with increase in offered load. Figure 5.13(a) and Figure 5.13(b) present only the real-time traffic.

It is seen from the above results that the proposed model provides QoS to limit packet loss and delay for each of the service from a real video conferencing node until the total offered load is up to the bottleneck limit of 2.1 Mbps. Thereafter at higher total offered load the AF and EF packets continue to get guarantee of low packet loss and delay whereas BE traffic gets affected. When total offered load exceeds the bottleneck bandwidth the priorities for AF and EF classes assure higher QoS for the classes repetitively where as the BE traffic class has higher packet loss and delay.

## 5.4.3 Performance of network with QoS traffic engineering for robotic surveillance camera

The snapshot of the quality of video transmission by robotic surveillance camera is shown in the Figure 5.14. There is a camera at one end and a web client at other end. The left side are the quality without proposed QoS model in the presence of total offered traffic load of 2.4 Mbps where the available bandwidth is 2.1 Mbps. The right side of the figure gives the clear picture with enabling proposed QoS and the presence of same offered traffic load. The absence of QoS shows the hazy and blurred images whereas in the presence of QoS the image is pretty clear.

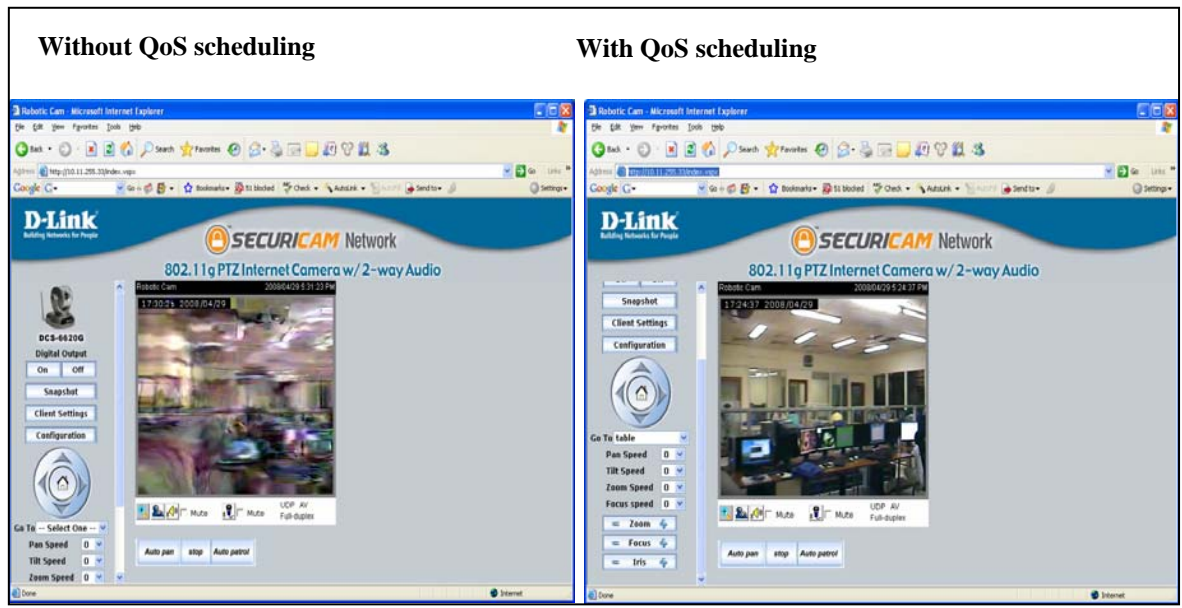

Figure 5.14: QoS performance improvement of real-time traffic using robotic surveillance camera as source

The performance statistics for video traffic are presented in Table 5.2.

Table 5.2: QoS performance statistics for robotic surveillance camera

| Traffic | Packet Loss | | Packet delay | |
|---|---|---|---|---|
| | With QoS | Without QoS | With QoS | Without QoS |
| Voice traffic | 0 % | 17 % | 110 ms | 216 ms |

## 5.4.4 Experiment with remote side on a real network

The experiment has also been conducted with remote side as receiver. The Tandberg video conferencing equipment is the real-time traffic generation tool and DITG is the cross traffic or traffic load generator. Figure 5.15 is the block diagram of the experiment setup for using remote side as receiver.

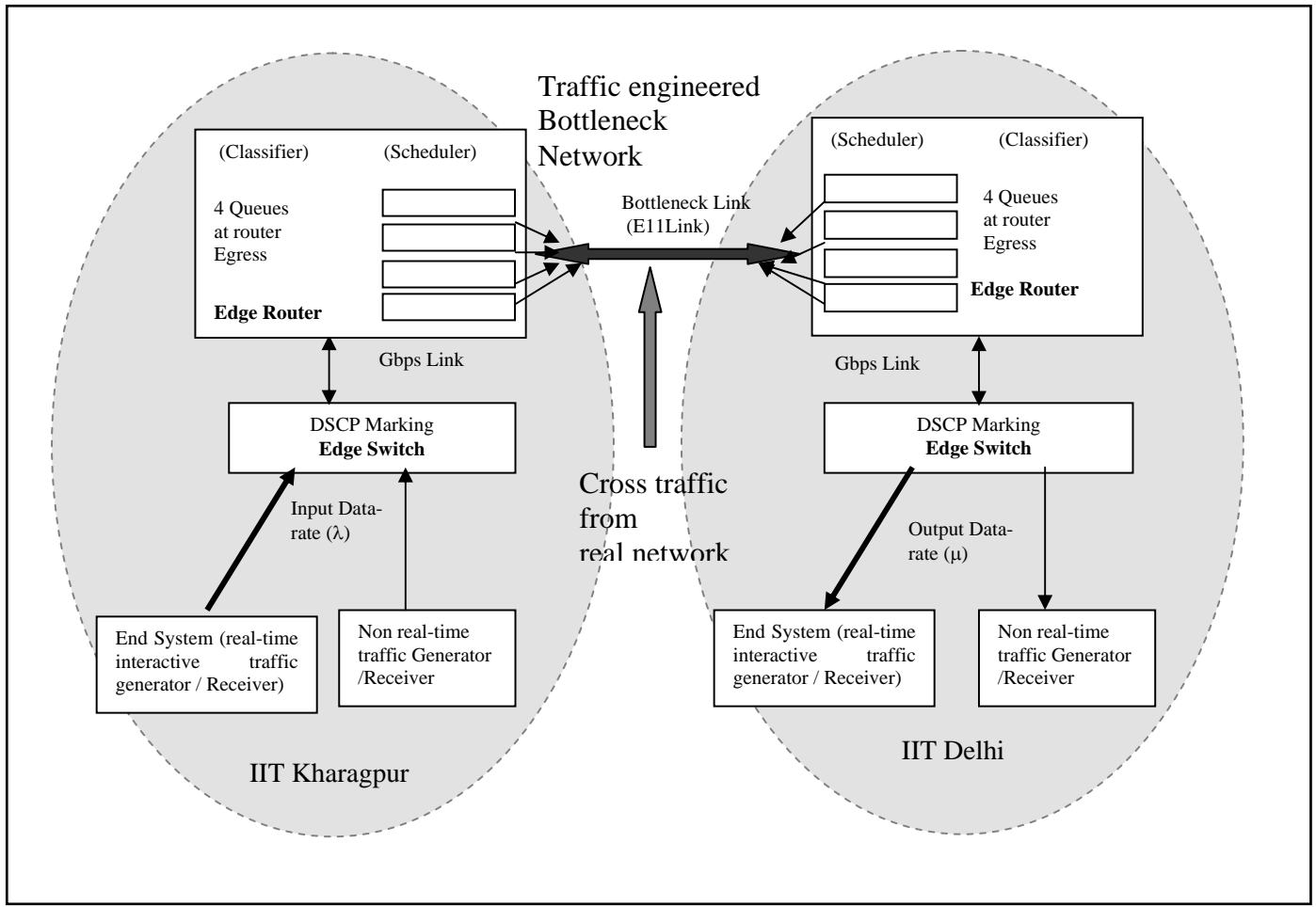

Figure 5.15: Test bed setup for QoS performance testing with endpoint at IIT Kharagpur and Remote-end point at IIT Delhi - Real test on E1

The total offered traffic load at the shared link is 2.6 Mbps whereas the link bandwidth is 2.1 Mbps. The performance statistics are presented at Table 5.3.

Table 5.3: IIT Delhi to IIT Kharagpur testing result (with 2 Mbps network)

| Traffic | Packet Loss | | Packet delay | |
|---|---|---|---|---|
| | With QoS | Without QoS | With QoS | Without QoS |
| **Voice traffic** | 0 % | 10 % | 140 ms | 180 ms |
| **Video traffic** | 0.9% | 21% | 20 ms | 35 ms |

It is quite clear from the several experimental results that the suggested QoS model assures the quality of real-time interactive traffic in the presence of high traffic load at the bottleneck link. The packet losses are almost negligible and the packet delays observed are not crossing the limiting value of performance requirements. It has been seen that the BE traffic are suffering from a bad quality which can also be adjusted by introducing some congestion control mechanism for TCP BE traffic. This has not been considered as it is beyond the scope of this thesis work.

## 5.5 Sensitivity analysis by real-time VBR service rate tuning parameter

The Real-Time VBR Service Rate Tuning Parameter, K was introduced in the previous chapter to control the service rate of the AF queue for video traffic [Section 4.4]. Here the experimental results for different K values have been also presented. The following graphs present the packet loss and delay for different value of K varying from 0.4 to 1.2. Figure-5.16 and Figure-5.17 show the % packet loss and packet delay respectively for VBR-rt traffic in the AF queue versus the normalized total offered load.

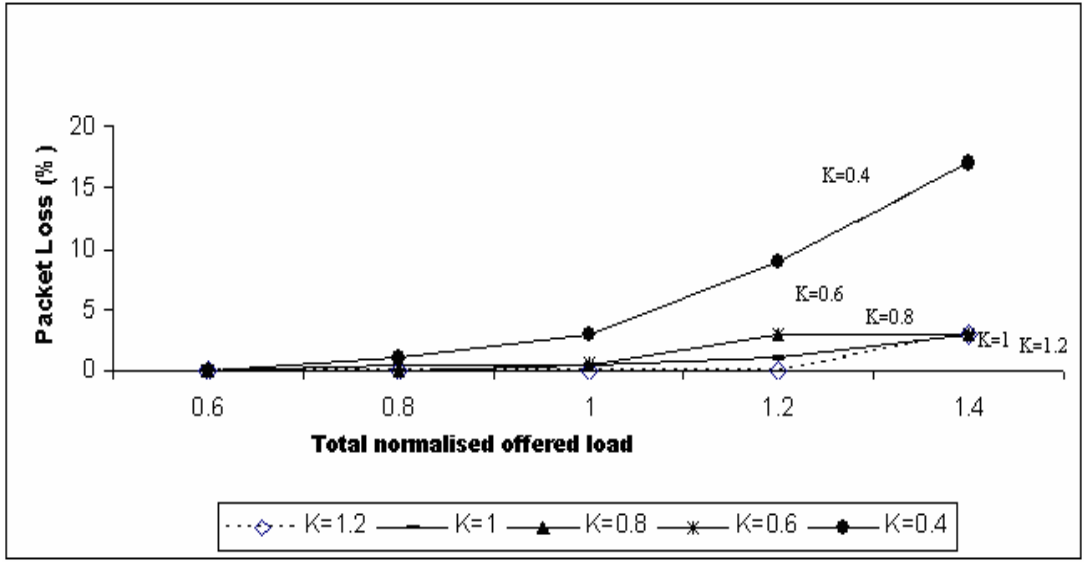

Figure 5.16: Packet loss vs. total normalized offered load for different service rate multiplying factor (K) for AF traffic

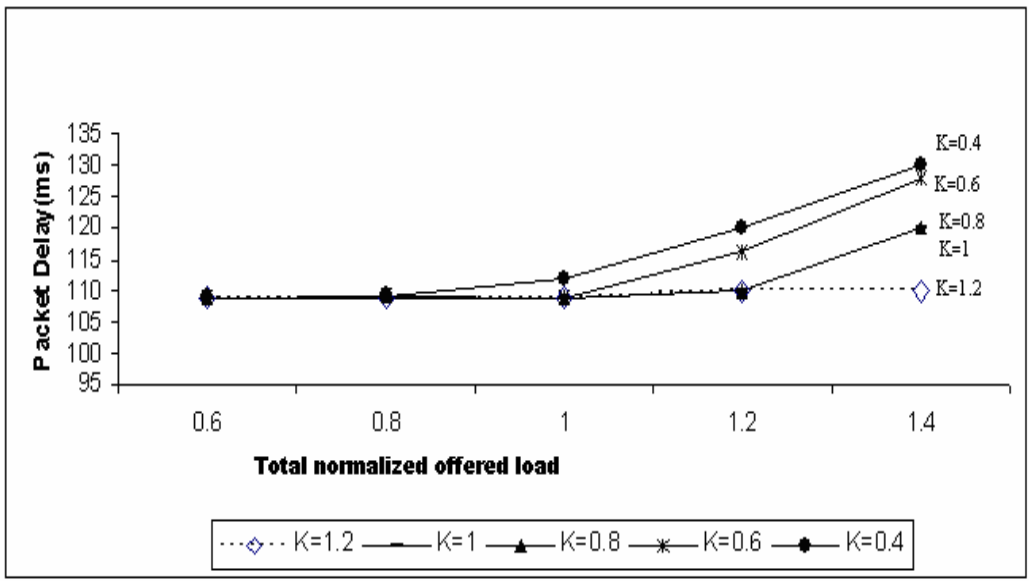

Figure 5.17: Packet delay vs. total normalized offered load for different service rate multiplying factor (K) for AF traffic

Figure 5.16 and Figure 5.17 show that for K = 0.6 the packet loss is 0% and delay < =120, for total offered load 2.1 Mbps. So when the total offered load is within available bandwidth to the link, the allocated service rate (queue weightage) for AF traffic can be brought down up to 3/5 times of the computed service rate, with same assured AF traffic quality of service. If the total offered load is increased beyond the available bandwidth, the allocated service rate of AF traffic can be brought down up to 4/5 times of the allocated service rate.

This sensitivity analysis now helps to decide how much variation of allocated service rate can be made without compromising the quality of the traffic. The advantage is that one can now take 1.6 times the no of VBR-rt sessions in the same computed AF service rate without reconfiguring the queues. Alternatively one can expect better performance to the other traffic even when such allocation is made.

All this shows that there is no need to over provision bandwidth at this level. Adequate tuning of the service can allow better usage of the available bandwidth. Reasonable QoS can be achieved by allowing a smaller allocation to real-time traffic and also keeping aside larger bandwidths for BE traffic.

# Chapter 6
# Conclusions

In this thesis work, a QoS model has been proposed to provide higher performance to real-time interactive traffic in constrained bandwidth networks. A DiffServ based approach has been proposed for use at the edge of a network with limited bandwidth resources and it ensures QoS and performance guarantees without needing bandwidth over provisioning to the users. The model is simulated with standard NS2 simulator to test and validate the proposed QoS scheduling algorithm. Finally the model has been implemented in a test bed by creating actual traffic engineered scenario. The proposed QoS model is of dynamic in nature as QoS controlling parameters are adaptive with varying queue length of different classes of traffic. Due to limitations of implementation in experimental setup the model has been slightly modified. In the experiment the parameters controlling the QoS are adaptive with input data rate for different classes of traffic. The model has been also tested with multiple domains of input sources sharing the same bottleneck link. This represents a typical network resource aggregation carried out at service provider levels. The simulation results show similar nature as with single domain. If load increases the packet loss and delay increase for the best effort traffic and at the same time assuring the quality of AF and EF traffic. From the simulation and experiment it is clear that proposed QoS mechanism improves the quality of the real-time traffic. The packet delay and packet loss are within the quality requirements of typical real time voice and video applications.

## 6.1 Contributions of the thesis work

In this thesis work, a method to tune the allocation of resources in a traffic engineering model to provide QoS, has been proposed to improve the performance of real-time traffic in a constrained bandwidth network. The model is formulated and there after simulated with a standard NS2 simulator to test and validate it.

Finally it has been implemented in an experimental setup emulating a typical real-time WAN scenario in the laboratory to create an actual traffic engineered scenario. The tests have been carried out on a router based test bed with modern day routers having routing and control planes segregated. The parameters considered in the study are buffer length and bandwidth ratios. This test-bed being primarily Juniper M7i router and Nortel 3510 switch based, the results presented here are specific to their implementation. However it can be studied in similar manner for other implementations also.

The QoS model has been tested with different source of real-time traffic having different voice and video codec. A Java based videoconferencing application test tool has been developed. It produces voice and video packets of different codes of different bit rates. Standard video conferencing system (Tandberg) and a robotic surveillance video camera have also been used for real-time traffic generation and verification of performance on a real network.

Experiments have also been conducted with a remote site (IIT Delhi Networks Laboratory) on a real network. The results shows that the QoS model improves the real-time traffic performance even in presence of heavy best effort cross traffic.

From the simulation and experiment it is clear that proposed QoS mechanism improves the quality of the real-time traffic in constrained bandwidth networks without need of over provisioning. The packet delay and packet loss are within the quality requirement of the application. The model has been verified with multiple domain of input source sharing the same bottleneck link.

The thesis also introduces a real-time VBR traffic tuning parameter for controlling the QoS services to the assured forwarding traffic so as to give better and fair performance to the rest of the traffic. The parameter may be used in admission control by controlling the number of session of video traffic. The tuning parameter is an index of resource allocation towards priority traffic as like the assured forwarding traffic typically used in real time interactive video. A lower than 1 value showing lesser allocation frees up more resources for best effort traffic. It is seen that even at values of 0.55 the parameter assures reasonably good performance for video. Such provisioning is useful for service providers to maintain network resources without over provisioning bandwidth for real time video and audio traffic. This can lead to saving on cost and energy for service providers.

## 6.2 Recommended implementation framework of the proposed QoS model

The proposed QoS model is dynamic in nature as QoS controlling parameters are also adaptive with varying queue length of different classes of traffic. Given the limitation of implementation in an experimental setup, the model has been slightly modified. Here the QoS controlling parameters are adaptive with input data rates of different classes of traffic. The suggested framework of proposed QoS mechanism using call admission control is shown in Figure- 6.1. Call admission control is not the scope of this thesis work. This work concentrates on adaptive QoS scheduling algorithm which has been implemented and tested successfully.

The QoS implementation framework is divided into two following parts such as-

A) <u>Call Admission Control procedure:</u>
1) The categorized traffic source interested in sending packets first consults with server agent to get permission for sending packets.
2) Serving agent then passes the request to monitoring agent.
3) Monitoring agent then asks the router for availability of resources.
4) The router replies the monitoring agent about the available resource.
5) The monitoring agent tells the serving agent whether the resource is available for a new call.
6) The serving agent sends a message to traffic source giving permission to send traffic if there is available resource or deny if resource is not available. The call admission control count for the number of sessions is updated for voice and video traffic.
7) Getting permission from serving agent traffic source starts sending packets.

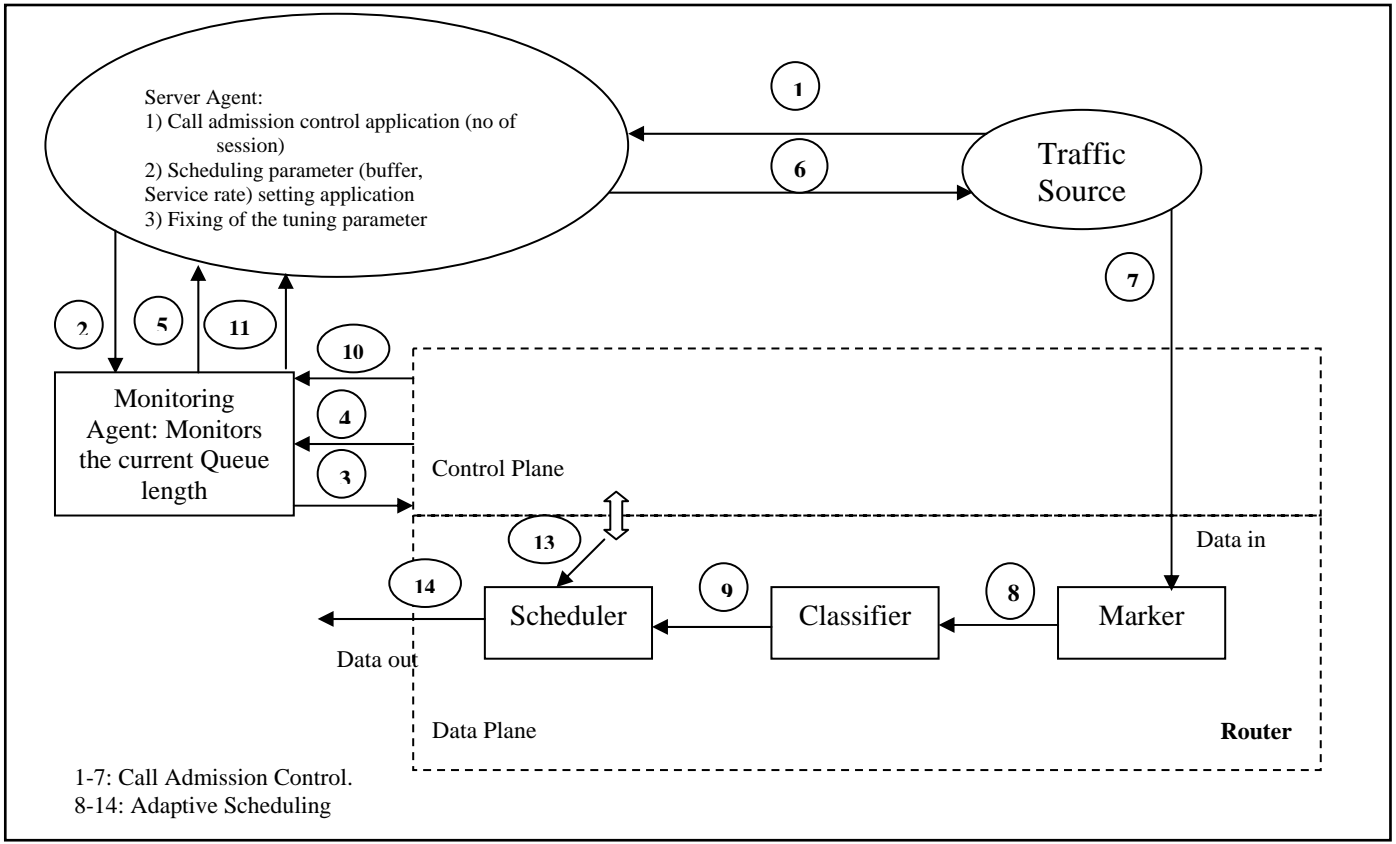

Figure 6.1: Recommended QoS implementation framework

B) Adaptive Scheduling procedure:
8) The packets coming from the traffic sources are marked with DSCP value for different real-time traffic.
9) The marked traffic are classified into several classes of services and distributed among the queues available.
10) The queues are scheduled with available bandwidth and buffer according to the scheduling algorithm discussed previously. A tuning parameter value is also taken when optimization is required.
11) The router ask serving agent about the input data rate of the traffic.
12) The serving agents provide router with the current input data rate of the traffic.

13) The router uses the information to calculate scheduling parameter to serve the traffic.
14) The packets are served to output data line.

The above QoS mechanism is adaptive with the current input rate of the traffic and available resource of the router. Here the router resources mean the buffer and bandwidth available at the router.

## 6.3 Scope of future work

The thesis work may be continued with various aspects. The scope of future work of work is as following:

- This thesis work concentrates on improving the QoS for real-time traffic. Best effort traffic performance can also be improved by implementing congestion control mechanism for TCP traffic, using RED model by developing similar strategies.
- For multiple domains of input sources, QoS scheduler can be implemented on the traffic for sharing the common bottleneck link. MPLS QoS mechanism also can be considered to improve the performance of the traffic.
- The QoS architecture for the router suggested is adaptive with the current state of the router and may be implemented by programming Monitoring agent and Server agent along with a Call Admission Control procedure.

# Appendix – I
# NS2 Simulator

Ns2 is a free network simulation program that can be downloaded from the web and is compatible with a number of operating systems [25]. The tool has substantial functionality for simulating different network topologies and traffic models. It also has an open architecture that allows users to add new functionality.

Ns2 has been developed at the Lawrence Berkeley National Laboratory (LBNL) of the University of California, Berkeley (UCB). DiffServ modules that were developed using ns-2.1b3 works on any update of ns-2, which is the current version of ns.

Ns2 is an event-driven network simulator. It has an extensible background engine implemented in C++ that uses OTcl (an object oriented version of Tcl) as the command and configuration interface. Thus, the entire software hierarchy is written in C++, with OTcl used as a front end.

## 1. Topology creation

Two nodes are connected by a link. The command for creating node is :

```
set n0 [$ns node]
set n1 [$ns node]
```

Command for creating link between nodes is :

```
$ns <link_type> $n0 $n1 <bandwidth> <delay> <queuetype>
$ns duplexlink $n0 $n1 1Mb 10ms dsRED/edge
```

## 2. Traffic generation

Command for creating UDP agent is :

```
set udp0 [new Agent/UDP]
$ns attach agent
$n0 $udp0
```

Create CBR traffic source for feeding into UDP agent:
```
set cbr0 [new Application/Traffic/CBR]
$cbr0 set packetSize_ 500
$cbr0 set interval_ 0.005
$cbr0 attach agent
$udp0
```
We can create Variable bit rate traffic (VBR) and variable packet size traffic with varying arrival rate and packet size with some distribution (exponential distribution, Poisson distribution etc.).

Traffic sink is created with the following command.
```
set null0 [new Agent/Null]
$ns attach agent
$n1 $null0
```

## 3. DiffServ implementation

In order to design and implement the DiffServ architecture in ns, five modules had to be added to the class hierarchy: one for the base DiffServ router functionality (dsRED), one each for the edge and core routers, one for RED-based queuing and one for policing. Each module defines a single class.

The dsRED module is the base module for the Diffserv implementation. It defines the *dsREDQueue* class, which is the parent class for the *edgeQueue* and *coreQueue* classes. The dsRED module is contained in the files "dsred.h" and "dsred.cc."

Following are some important codes used to implement DiffServ QoS in the simulation.
```
    # The number of physical queues
     $q($a$b) set numQueues_ 3
    # Set scheduler (and queue weigths)
     $q($a$b) setSchedularMode WRR ;
     $q($a$b) addQueueWeights 0 $EF(qw) ;
     $q($a$b) addQueueWeights 1 $AF(qw)
     $q($a$b) addQueueWeights 2 $BE(qw)
```

```
 # Configure PHBs
   # EF
   $q($a$b) addPHBEntry $EF(cp) 0 0 ;
   $q($a$b) set PhysQueueSize_ 0 $EF(qlimit)
   # AF
   $q($a$b) addPHBEntry $AF(cp) 1 0 ;
   $q($a$b) set PhysQueueSize_ 1 $AF(qlimit)
   #BE
   $q($a$b) addPHBEntry $BE(cp) 2 0 ;
   $q($a$b) set PhysQueueSize_ 2 $BE(qlimit)
```

## 4. Monitoring

Packet loss and delay can be viewed by parsing .tr file generated during the simulation [26]. .tr file has fields with following meaning:

```
  action = $1;
  time = $2;
  from = $3;
  to = $4;
  type = $5;
  pktsize = $6;
  flow_id = $8;
  src = $9;
  dst = $10;
  seq_no = $11;
  packet_id = $12;
```

# Appendix – II
# Traffic Generator

Distributed Internet Traffic generator (DITG) is used to generate Best Effort traffic for experimental study of the thesis work. Following commands are used to generate and receive traffic [30].

**ITGRecv:** It receives flows from different senders. By this command the receiver opens a port and waits for data sent by the user.

**ITGSend:** It generates flows of traffic. Following options are used in the experiments.

| Flow Options | Meaning |
| --- | --- |
| –a <destination address> | Set the destination address. Default is local host. |
| –T <protocol type> | Set the protocol type. Valid values are UDP, TCP, ICMP, SCTP, and DCCP. Default is UDP. If you choose ICMP you must specify the type of message. Root privileges are needed under Linux. |
| –t <duration> | Set the generation duration. It is expressed in milliseconds. Default is 10000 ms. |
| **Inter Departure time options :** | |
| –C <pkts_per_s> | Constant inter-departure time (IDT). |
| –U <min_pkts_per_s> <max_pkts_per_s> | Uniformly distributed IDT. |
| –E <average_pkts_per_s> | Exponentially distributed IDT. |
| –V <shape> <scale> | Pareto distributed IDT. |
| –Y <shape> <scale> | Cauchy distributed IDT. |
| –N <mean> <std dev> | Normal distributed IDT. |
| –O <average pkts_per_s> | Poisson distributed IDT. |
| –G <shape> <scale> | Gamma distributed IDT. |
| **Packet size options:** | |
| –c <pkts_size> | Constant payload size. |
| –u <min_pkts_size> <max_pkts_size> | Uniformly distributed payload size. |
| –e <average_pkts_size> | Exponentially distributed payload size. |
| –v <shape> <scale> | Pareto distributed payload size. |
| –y <shape> <scale> | Cauchy distributed payload size. |
| –n <mean> <std_dev> | Normal distributed payload size. |
| –o <average_pkts_size> | Poisson distributed payload size. |
| –g <shape> <scale> | Gamma distributed payload size. |

# Appendix – III
# Juniper Class of Service Configuration

Class of Service (CoS) is the assignment of traffic flows to different service levels that provide different latency, jitter and loss characteristics to particular applications served by specific traffic flow.

JUNOS CoS functionality is made possible through a series of mechanisms as shown in Figure – 1.

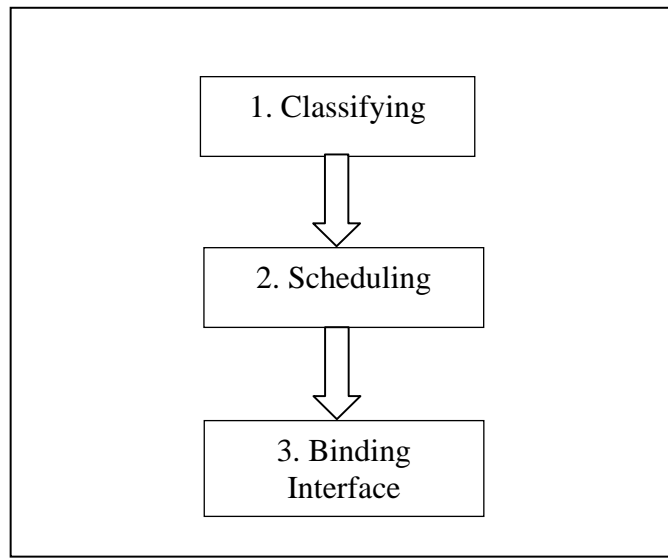

Figure 1. JUNOS CoS Mechanism

Classifiers aggregate different traffic types identified into a forwarding class by using IP precedence, DSCP or IEEE 802.1p user precedence bit or by MPLS EXP field. After a forwarding class assigned to the transmission queue, there is a one to one correspondence between the queues and forwarding classes. Schedulers are assigned to the forwarding classes by use of a scheduler map. The scheduler map is then assigned to an interface. The scheduler defines the queuing parameters such as transmission rate, buffer size and priority. Following is the sample CoS configuration for Junos 7.5 [33].

1. <u>Classifying into different forwarding classes</u>

```
classifiers {
    dscp QoS-testbed {

forwarding-class expedited-forwarding {
                    loss-priority low code-points 101110;
        }

forwarding-class assured-forwarding {
                    loss-priority low code-points 001100;
        }
forwarding-class best-effort {
                    loss-priority high code-points 000000;
        }
forwarding-class network-control {
                    loss-priority low code-points 110000;
        }
    }
}
forwarding-classes {
    queue 0 best-effort;
    queue 1 expedited-forwarding;
    queue 2 assured-forwarding;
    queue 3 network-control;
}
```

2. Configuring Scheduler

```
scheduler-maps {
    QoS {
     forwarding-class assured-forwarding scheduler video;
     forwarding-class expedited-forwarding scheduler voice;
     forwarding-class best-effort scheduler other-traffic;
     forwarding-class network-control scheduler network;
    }
}
schedulers {
video {
        transmit-rate percent 20 ;
        buffer-size percent 40;
        priority high;
    }
voice {
        transmit-rate percent 5;
        buffer-size percent 10;
        priority strict high;
    }
network {
         transmit-rate percent 5;
         buffer-size percent 5;
         priority high;
    }
other-traffic {
        transmit-rate remainder;
        buffer-size remainder;
        priority low;
    }
}
```

3. <u>Binding Scheduler with interfaces</u>

```
interfaces {
   e3-0/1/1 {
        scheduler-map QoS;
        unit 0 {
            classifiers {
                dscp QoS-testbed;
            }
        }
    }
   ge-1/3/0 {
        scheduler-map QoS;
        unit 0 {
            classifiers {
                dscp QoS-testbed;
            }
        }
    }
   e1-0/2/1 {
        scheduler-map QoS;
        unit 0 {
            classifiers {
                dscp QoS-testbed;
            }
        }
    }
}
```